\documentclass[runningheads]{llncs}
\usepackage[T1]{fontenc}
\usepackage[ruled, linesnumbered, vlined]{algorithm2e}
\usepackage{graphicx}
\usepackage{amsmath}
\usepackage{amssymb}
\usepackage{wrapfig}

\newtheorem{defn}{Definition}
\newcommand{\nonl}{\renewcommand{\nl}{\let\nl\oldnl}}
\begin{document}
\title{A Pragmatic Approach to \\Stateful Partial Order Reduction}
%
%
\author{Berk Cirisci\inst{1}\orcidID{0000-0003-4261-090X} \and
Constantin Enea\inst{2}\orcidID{0000-0003-2727-8865} \and
Azadeh Farzan\inst{3}\orcidID{0000-0001-9005-2653} \and
Suha Orhun Mutluergil\inst{4}\orcidID{0000-0002-0734-7969}}

\authorrunning{B. Cirisci et al.}
%
\institute{IRIF, Universit\'e Paris Cit\'e \\
\email{cirisci@irif.fr} \and
LIX, Ecole Polytechnique, CNRS and Institut Polytechnique de Paris \\
\email{cenea@lix.polytechnique.fr} \and
University of Toronto \\
\email{azadeh@cs.toronto.edu} \and
Sabanci University \\
\email{suha.mutluergil@sabanciuniv.edu}}
\maketitle              

\begin{abstract}
Partial order reduction (POR) is a classic technique for dealing with the state explosion problem in model checking of concurrent programs. Theoretical optimality, i.e., avoiding enumerating equivalent interleavings, does not necessarily guarantee optimal overall performance of the model checking algorithm. The computational overhead required to guarantee optimality may by far cancel out any benefits that an algorithm may have from exploring a smaller state space of interleavings. With a focus on overall performance, we propose new algorithms for stateful POR based on the recently proposed source sets, which are less precise but more efficient than the state of the art in practice. We evaluate efficiency using an implementation that extends Java Pathfinder in the context of verifying concurrent data structures.
\end{abstract}
\def\ample{{\sc S-POR}}
\def\eager{{\sc DE-S-POR}}
\def\lazy{{\sc DL-S-POR}}


\section{Introduction}

Concurrency results in insidious programming errors that are difficult to reproduce, locate, and fix. Therefore, verification techniques that can automatically detect and pinpoint errors in concurrent programs are invaluable. 
\emph{Model checking}~\cite{DBLP:conf/popl/ClarkeES83,DBLP:conf/programm/QueilleS82} explores the state space of a given program in a systematic manner and verifies that each reachable state satisfies a given property. It provides high coverage of program behavior, but it faces the infamous state explosion problem, i.e., the number of possible thread interleavings grows exponentially in the size of the source code. In this paper, we consider shared-memory programs running on a sequentially consistent memory model, for which interleavings of atomic steps in different threads are a precise model of concrete executions.

\vspace{-.2mm}
\emph{Partial order reduction} (POR)~\cite{DBLP:journals/sttt/ClarkeGMP99,DBLP:books/sp/Godefroid96,DBLP:conf/cav/Peled93,DBLP:conf/apn/Valmari89} is an approach that limits the number of explored interleavings without sacrificing coverage. POR relies on an equivalence relation between interleavings, where two interleavings are equivalent if one can be obtained from the other by swapping consecutive independent (non-conflicting) execution steps. It guarantees that at least one interleaving from each equivalence class (called a Mazurkiewicz trace~\cite{DBLP:conf/ac/Mazurkiewicz86}) is explored. Optimal POR techniques explore exactly one interleaving from each equivalence class. Beyond this classic notion of optimality, POR techniques may aim for optimality by avoiding visiting states from which no optimal execution may pass. There is a large body of work on POR techniques that address its soundness when checking a certain class of specifications for a certain class of programs, or its theoretical optimality (see Section~\ref{sec:related}). The set of interleavings explored by some POR technique is defined by restricting the set of threads that are explored from each state (scheduling point). Depending on the class specifications, assumptions about programs, or optimality targets, there are various definitions for this set of processes, including stubborn sets~\cite{DBLP:conf/apn/Valmari89}, persistent sets~\cite{DBLP:books/sp/Godefroid96}, ample sets~\cite{DBLP:journals/sttt/ClarkeGMP99}, and source sets~\cite{DBLP:journals/jacm/AbdullaAJS17}.

\vspace{-.2mm}
The design of a model checking algorithm based on POR has to consider several computational tradeoffs. First, such an algorithm can be stateful or stateless~\cite{DBLP:conf/popl/Godefroid97}, which corresponds to a tradeoff between memory consumption versus execution time. Stateful model checking records visited states, thereby consuming more memory, but stateless model checking performs redundant exploration from already visited states. Second, the computation of the set of threads that are explored from some state can be more or less complex. Focusing on theoretical optimality, e.g., exploring \emph{exactly} one interleaving from each Mazurkiewicz trace, may make this computation more complex. This complexity in turn may diminish the overall performance when the potential for reducing the state space is not large, i.e., most Mazurkiewicz traces contain a small number of interleavings. In such a case, exploring more interleavings can take less time than computing more precise constraints on the explored schedules. Third, POR algorithms may compute the information they use for the purpose of reduction {\em statically}, by some kind of conservative static analysis of the source code,  or {\em dynamically}, during the exploration of interleavings. Static computation is usually cheaper and less precise than dynamic computation. 

\vspace{-.2mm}
In this work, we investigate the use of POR from a practical point of view.  In the context of verifying concurrent data structures, we investigate the following research question: what tradeoffs in POR families of algorithms may lead to practical net gains in verification or bug-finding times? We focus on the application domain of verification of Java concurrent data structures using a tool like Violat~\cite{DBLP:conf/cav/EmmiE19}. Concurrent data structures provide implementations of common abstract data types (ADTs) like queues, key-value stores, and sets. 
Their correctness amounts to observational refinement~\cite{DBLP:conf/esop/HeHS86,DBLP:journals/ipl/HoareHS87,DBLP:journals/tcs/Plotkin77} which captures the substitutability of an ADT with an implementation~\cite{DBLP:journals/toplas/LiskovW94}: any combination of values admitted by a given implementation is also admitted by the given ADT specification. Violat can be used to generate tests of observational refinement, i.e., bounded-size clients of the concurrent data structure that include assertions to check that any combination of return values observed in an execution belongs to a statically precomputed set of ADT-admitted return-value outcomes. Violat is integrated with the Java Pathfinder (JPF) model checker~\cite{DBLP:conf/issta/VisserPK04}, which enables complete systematic coverage of a given test program and outputting execution traces leading to consistency violations, thus facilitating diagnosis and repair. We investigate POR algorithms implementable in JPF. 

\vspace{-.5mm}
We study several stateful model checking algorithms with POR in the context of Violat's test programs. This choice was inspired by experiments that demonstrated that it is much faster than the stateless variation (see Section~\ref{sec:exp}). We introduce POR algorithms that combine \emph{static and dynamic} computations of sets of threads to explore from a given state. In the context of stateful model checking, static techniques may seem like the better option. A dynamic computation usually requires re-traversing the state space starting in an already visited state which can be time consuming.  Note however that re-traversing the state space that is already loaded in memory takes less time than generating that state space in the first place, which involves executing program statements.

\vspace{-.5mm}
Our starting point is a simple static POR algorithm, called {\ample}, that makes use of  \emph{invisible} transitions. These transitions are independent of any transition of another concurrently-executing thread (they correspond to the safe actions introduced in~\cite{DBLP:conf/forte/HolzmannP94}). Based on a syntactic analysis of the code, we identify shared and synchronization objects, and assume that every transition that does not access such an object is invisible. For clients of concurrent data structures, such objects correspond to class fields accessed in a method of the data structure. Invisible actions include starting and joining threads, and method calls and returns, for instance. The POR algorithm prioritizes the exploration of invisible transitions over visible ones, i.e., if an invisible transition is enabled in a given state then this is the only explored transition from that state, and otherwise, all enabled transitions are explored. 
We demonstrate that {\ample} has a small overhead and the potential for substantial reductions, and therefore leads to significant speedups with respect to standard JPF which employs a very conservative heuristic for its POR (see Section~\ref{sec:exp}).

\vspace{-.5mm}
{\ample} is effective, but by the nature of being lightweight, does not always reduce the state space effectively. We introduce two new algorithms as extensions of {\ample}, with the idea of performing a more aggressive reduction while keeping the overhead reasonably low. They {\em dynamically} compute \emph{source sets},  which restrict the set of threads explored from a state with only {\em visible} enabled transitions. We focus on source sets since they are provably minimal, i.e., the set of explored threads from some state must be a source set in order to guarantee exploration of all Mazurkiewicz traces. Moreover, any superset of a source set is also a source set, which makes their computation less sensitive to the order in which transitions from a given state are enumerated. This property does not hold for other definitions such as stubborn sets, persistent sets, or ample sets.

\vspace{-.5mm}
The design principle behind our algorithms is to favor efficiency over theoretical optimality. 
Our algorithms are not theoretically optimal. However, we demonstrate that they are more efficient than the optimal algorithm~\cite{DBLP:journals/jacm/AbdullaAJS17} where the overhead of source set computation subsumes any gains from not exploring the redundant interleavings. 

\vspace{-.5mm}
In general, a dynamic computation of source sets relies on tracking dependencies between actions in the already explored executions. Our two proposed algorithms differ in the way in which the tracking is performed: one is \emph{eager} and called {\eager},  and the other is \emph{lazy} and called {\lazy}.  
Intuitively, {\eager} advances the computation of source sets for predecessors in the current execution in a style similar previous dynamic POR algorithms, e.g.~\cite{DBLP:conf/popl/FlanaganG05,DBLP:journals/jacm/AbdullaAJS17}, while {\lazy} advances the computation of the source set in a given state only when the exploration backtracks to that state and one must decide if a new transition has to be explored.  

The thesis of this paper is that  when there is a big enough potential for reducing the state space of a concurrent program, i.e., many Mazurkiewicz traces are large enough, non-optimal but carefully customized algorithms, like {\eager} and {\lazy}, can have the largest impact compared to the two extremes of the spectrum, that is, {\ample} or theoretically optimal algorithms like~\cite{DBLP:journals/jacm/AbdullaAJS17}. If the potential for reduction is small, then a simple static algorithm like {\ample} provides the best overhead-gain tradeoff. 

To support this thesis, we implemented these algorithms in JPF and evaluated them on a number of clients of concurrent data structures from the Synchrobench repository~\cite{DBLP:conf/ppopp/Gramoli15}. Our evaluation shows that they outperform (1) their variations that are directly built on top of the standard setup of JPF, (2) their stateless variations, and (3) a best-effort implementation of a stateful variation of the optimal algorithm in~\cite{DBLP:journals/jacm/AbdullaAJS17} . The lazy algorithm {\lazy} is more efficient than the eager {\eager}, and more efficient than {\ample} on clients with a big enough potential for reducing the state space. 
\section{Preliminaries}

We model a concurrent (multi-threaded) program with a bounded number of threads as a labeled transition system (LTS) $L = (\mathcal{S}, s_I, \Gamma)$. We assume that programs run under sequential consistency. A state in $\mathcal{S}$  represents a finite set of \emph{shared} objects visible to all threads and a finite set of \emph{local} objects visible to a single fixed thread, and a program counter for each  thread. The state $s_I\in \mathcal{S}$ is the unique initial state. $\Gamma$ is a set of labeled transitions $(s,a,s')$ where $s,s'\in\mathcal{S}$ and $a$ is an \emph{action} (transition label) representing the execution of an \emph{atomic} statement in the code. Action $a$ records the executing thread, its program counter, and shared object accesses. There are two types of actions: (1) \emph{invisible actions}: $a=(t,\mathit{pc}, \epsilon)$ where a thread $t$ executes a statement at program counter $\mathit{pc}$ that accesses no shared object, and (2) \emph{visible actions}: $a=(t,\mathit{pc},r/w,o)$ where $t$ executes a statement at $\mathit{pc}$ that reads ($r$) or writes ($w$) the shared object $o$.
For an action $a$, $\mathit{tid}(a)$ is the thread id $t$, and $\mathit{op}(a)$ and $\mathit{obj}(a)$ refer to the third and fourth components when $a$ is visible (otherwise they are undefined). 

A transition labeled by a visible (resp. invisible) action is called visible (resp. invisible). In the context of a full-fledged programming language, invisible transitions are related to local computations, control-flow manipulations (e.g., starting/stopping threads and calling or returning from a method), or accesses to ``low-level'' shared objects that are irrelevant for the intended (functional) specification. Visible transitions correspond to the execution of a single atomic statement that accesses a shared object followed by a maximal sequence of \emph{local} statements that only modify the local states of that thread.

We assume that LTSs are deterministic and acyclic. 
An action $a$ is \emph{enabled} in state $s$ if there exists $s'$ such that $(s,a,s')\in\Gamma$.
We use $\mathit{next}(s,t)$ to denote the transition $(s,a,s')\in \Gamma$ for some $a$ and $s'$ with $\mathit{tid}(a)=t$, if it exists, and $\mathit{succ}(s,t)$ to denote the successor $s'$ in this transition. Otherwise, we say that $t$ is \emph{blocked} in $s$. The set $\mathit{enabled}(s)$ is the set of threads that are not blocked in $s$. A state $s$ is \emph{final} if $\mathit{enabled}(s) = \emptyset$.
Two actions $a$ and $a'$ of different threads 
are \emph{independent} if they are both enabled in a state $s$ and either one of them is an invisible action, or they are both visible and access different shared objects ($\mathit{obj}(a)\neq \mathit{obj}(a')$), or they both perform a read access ($\mathit{op}(a)=\mathit{op}(a')=r$). The actions $a$ and $a'$ are called \emph{dependent}, denoted by $a \nsim a'$, if they are not independent. We assume that if an action $a$ enables or disables another action $a'$, then $a \nsim a'$. Two transitions are (in)dependent iff they contain actions that are (in)dependent.


An \emph{execution from a state $s$} is a sequence of alternating states and actions $E=s_0,a_0,s_1,a_1,\ldots,s_n$ with $s_0=s$ and $(s_i,a_i,s_{i+1})\in\Gamma$ for each $0\leq i\leq n-1$. The set of execution starting from $s$ in the LTS $L$ is denoted by $E(L,s)$.
An \emph{initialized} execution is an execution from $s_I$. Initialized executions that end with a final state are called \emph{full} executions.
We assume absence of deadlocks, i.e., a full execution $E$ contains every action enabled in a state of $E$. 

The \emph{happens-before} relation in an execution $E$, denoted by $\rightarrow_E$, captures the causal relation among actions in $E$ (the program order between actions of the same thread and the order between actions accessing the same shared object where at least one of them is a write). 
Given two actions $a$ and $a'$ labeling transitions in $E$, $a \rightarrow_E a'$ holds iff $a \nsim a'$ and the transition labeled by $a$ occurs before the transition labeled by $a'$ in $E$. Two executions $E$ and $E'$ are called \emph{equivalent} if $\rightarrow_E=\rightarrow_{E'}$. For a full execution $E$, we use $[E]$ to denote the set of full executions $E'$ that are equivalent to $E$.

Given an LTS $L=(\mathcal{S}, s_I, \Gamma)$ that models a concurrent program, an LTS $L_r=(\mathcal{S}_r, s_I, \Gamma_r)$ with $\mathcal{S}_r\subseteq \mathcal{S}$ and $\Gamma_r\subseteq \Gamma$ is called \emph{sound for $L$} if for each full execution $E$ of $L$, there exists a full execution $E'$ of $L_r$ that is equivalent to $E$. 

%
%
\vspace{-4mm}
\subsection{Partial Order Reduction}\label{ssec:por}

The set of executions explored by POR techniques is defined by restricting the set of threads that are explored from each state. The algorithms discussed in this paper fall into two categories in this respect: persistent sets and source sets. Both  guarantee soundness, i.e.,  at least one execution from each equivalence class is explored.

Intuitively, a set of threads $T$ is \emph{persistent} for a state $s$ if in any execution starting from $s$, the first transition that is dependent on some transition starting from $s$ of some thread $t\in T$ is taken by some thread $t'\in T$ ($t$ and $t'$ may be equal). A set of threads $T$ is a \emph{source} set for $s$ if for any execution starting from $s$, there is some thread in $T$ that can take the first step, modulo reorderings of independent transitions. We define persistent and source sets as sets of threads, which correspond to sets of transitions in the classical sense, under the assumption of determinacy of individual threads. 

\vspace{-2mm}
\begin{defn}[Persistent Set~\cite{DBLP:books/sp/Godefroid96}]
A set of threads $T$ is called a \emph{persistent set for a state $s$} if for every execution $E$ from $s$ that contains only transitions from thread $t' \not\in T$, every transition in $E$ is independent of every transition $\mathit{next}(s,t)$ with $t\in T$.
\end{defn}

%
%
\vspace{-3mm}
For an execution $E$ from a state $s$ that ends in a final state, a thread $t$ is called a \emph{weak initial} of $E$ if there exists an execution $E'$ that is equivalent to $E$ and starts with a transition of $t$.

\vspace{-2mm}
\begin{defn}[Source Set~\cite{DBLP:journals/jacm/AbdullaAJS17}]
A set of threads $T$ is called a \emph{source set for a state $s$} if every execution from $s$ that ends with a final state has a weak initial thread in $T$. 
\end{defn}

\vspace{-3mm}
An exploration where each state is expanded w.r.t. the threads in a persistent or source set is sound (when finished it produces an LTS which is sound for the ``full'' LTS of the program). However, source sets guarantee a stronger notion of optimality~\cite{DBLP:journals/jacm/AbdullaAJS17}. There exist programs where any persistent set (for the initial state) is strictly larger than a source set~\cite{DBLP:conf/birthday/AbdullaAJS17}, but every persistent set is also a source set. Note that source sets are monotonic in the sense that any superset of a source set is also a source set, but this is not true for other definitions such as stubborn sets, persistent sets, or ample sets.
\vspace{-2mm}
\section{Eager Source Set POR (DE-S-POR)}\label{sec:eager}


%
%

We present a first stateful POR algorithm that selects a sufficient set of threads to expand a state based on two criteria: (1) a static criterion based on (in)visible actions, and (2) a dynamic criterion based on \emph{source sets} computed on-the-fly during the exploration. Source sets are maintained \emph{eagerly} for each new transition that is explored, in a style similar to previous algorithms, e.g.~\cite{DBLP:conf/popl/FlanaganG05,DBLP:journals/jacm/AbdullaAJS17}. For presentation reasons, we start with a simplified version that includes only the static criterion and continue with the full version afterwards.
\vspace{-2mm}
\subsection{Safe Set POR (S-POR)}

\begin{figure}[h]
\begin{algorithm}[H]
\SetAlCapNameFnt{\scshape}
\SetAlgoVlined
\caption{Safe Set POR ({\ample})}
\label{alg:ample}
\SetKwInput{init}{Initialize}{}{}
\SetKwProg{search}{Explore}{()}{end}
\init{$Stack \gets \emptyset$; $Stack.\mathbf{push}(s_I)$; $L_r \gets \emptyset$;}
\search{}{
	$s \gets Stack.\mathbf{top}$;\\
	\If {$\mathit{notVisited}(s)$\label{line:stop}} {
		\ForAll {$t \in \mathit{safeSet}(s)$} {
			$(s,a,s') \gets \mathit{next}(s, t)$; \\
			$Stack.\mathbf{push}(s')$; \tcp{transition $(s, a, s')$ is added to $L_r$} 
			$\mathbf{Explore}()$;\\
			$Stack.\mathbf{pop}()$;
		}
	}
}
\nonl ${notVisited}(s)$ holds if $s$ is final in $L_r$ but $\mathit{enabled}(s)\neq\emptyset$
\[
	\hspace{-0.5cm} 
	\mathit{safeSet}(s)=
	\begin{cases}
		\{t\}, & \exists t \in \mathit{enabled}(s): \mbox{$\mathit{next}(s,t)=(s,a,s')$ and $a$ is invisible}\\
		\mathit{enabled}(s), & \mbox{otherwise}
	\end{cases}
\]
\end{algorithm}
\vspace{-0.8cm}
\end{figure}

\begin{wrapfigure}{r}{0.3\textwidth}
\vspace{-.5cm}
  \begin{center}
    \includegraphics[scale=0.55]{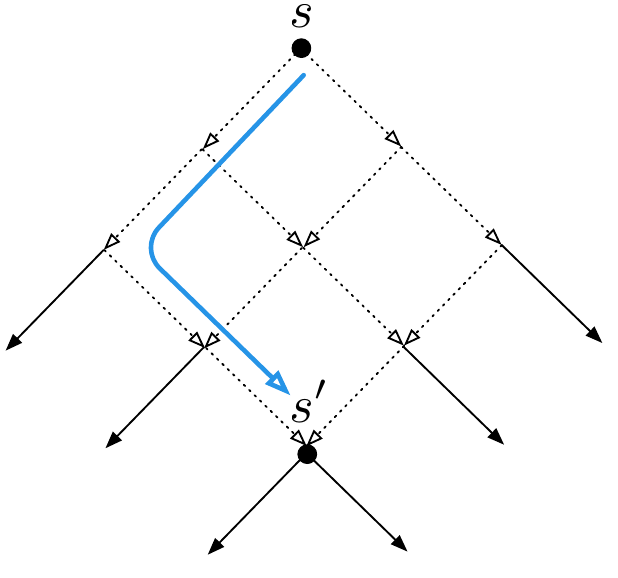}
  \end{center}
 \vspace{-24pt}
  \caption{Full traversal vs. partial {\ample} (in blue).}\label{fig:ample}
 \vspace{-10pt}
\end{wrapfigure}

Algorithm~\ref{alg:ample} presents a stateful DFS traversal of a concurrent program, represented by an LTS, which restricts the traversal to so called {\em safe sets}. Figure \ref{fig:ample} illustrates the core idea of this algorithm. The safe sets prioritize the exploration of {\em invisible} transitions over visible ones. 

For a state $s$, if there is an enabled thread $t\in\mathit{enabled}(s)$ whose enabled transition is invisible, then $\emph{safeSet}(s) = \{t\}$. Otherwise, $\emph{safeSet}(s)$ contains all the threads enabled in $s$, and $s$ is called an \textit{irreducible state}.  In Figure \ref{fig:ample}, only state $s'$ is irreducible since any other state has at least one enabled invisible transition, and all other states are reducible. 

In Algorithm~\ref{alg:ample}, $\mathit{Stack}$ represents the stack of the DFS traversal and it is considered to be a global variable, and $L_r$ records transitions explored during the traversal. Note that the DFS traversal stops the exploration whenever it visits a state $s$ that has been visited in the past (see the condition at line~\ref{line:stop}). The choice of safe sets then provides additional savings on top of the standard DFS traversal strategy. When the traversal ends, $L_r$ is sound (for the ``full'' LTS of the program). 



Observe that Algorithm~\ref{alg:ample} can reduce the number of visited states in a significant way. The diagram in Figure~\ref{fig:ample} corresponds to a fully explored program LTS while the path marked by the blue arrow is the result of Algorithm~\ref{alg:ample}. It is easy to observe that one can obtain an exponential reduction (with the base of the number of consecutive invisible transitions and the exponent of the number of threads) with this algorithm. 
 

\vspace{-3mm}
\subsection{Full Algorithm}

\begin{figure}[t]
\begin{algorithm}[H]
\SetAlCapNameFnt{\scshape}
\SetAlgoVlined
\caption{Eager Source Set POR ({\eager})}
\label{alg:eager}
\SetKwInput{init}{Initialize}{}{}
\SetKwProg{search}{Explore}{()}{end}
\SetKwProg{updateCurrent}{UpdateCurr}{($a$)}{end}
\SetKwProg{ampleSet}{safeSet}{(s)}{end}
\SetKwProg{updateBacktrack}{UpdateBacktrack}{(s, a)}{end}
\SetKwProg{isDone}{IsDone}{(s)}{end}

\init{$Stack \gets \emptyset$; $Stack.\mathbf{push}(s_I)$; $L_r \gets \emptyset$;}

\search{}{
	$s \gets Stack.\mathbf{top}$;\\
	\uIf {$\mathit{notVisited}(s)$} {
		\uIf {$\exists t \in \mathit{safeSet}(s)$ \label{line:start_unvis}} {
		$s.\mathbf{backtrack} \gets \{t\}$; $s.\mathbf{current} \gets \emptyset$; $s.\mathbf{done} \gets \emptyset$;\\
		\While {$\exists t'\in s.\mathbf{backtrack} \setminus s.\mathbf{done}$ \label{line:backEqDone}} {
			$(s,a,s') = \mathit{next}(s,t')$;\\
			$Stack.\mathbf{push}(s')$;\\
			$s.\mathbf{done}$ = $s.\mathbf{done}\cup \{t'\}$;\\
			$s.\mathbf{current}[t'] \gets \{t'\}$;\\
			\lIf {$a$ is visible} {$\mathbf{UpdateCurr}(a)$ \label{line:curr}}
			$\mathbf{Explore}()$;\\
			$s.\mathbf{backtrack} \gets \mathit{UpdateBack}(s, a)$; \label{line:updateBack}\\
			$Stack.\mathbf{pop}()$;\label{line:end_unvis}\\
		}
		}
	} \Else {
		$A_{s} \gets \{a' : \mbox{$a'$ occurs in an execution from $E(L_r,s)$} \}$;\label{line:start_revisit} \\
		\lForEach {$a' \in A_{s}$} {$\mathbf{UpdateCurr}(a' )$} \label{line:end_revisit} 
	}
}

\updateCurrent{}{
	$E$ is the initialized execution of $L_r$ following states in $Stack$;\\
	$(s,a',s')$ is the last transition of $E$ with $\hspace{-.3mm} a \hspace{-.6mm} \label{line:depExists}
	\nsim\hspace{-.6mm} a'\hspace{-1mm}\land\hspace{-.6mm} \mathit{tid}(a)\hspace{-.7mm}\neq\hspace{-.7mm}\mathit{tid}(a')$\\
	\If {$(s,a',s') \neq null$}{
		$s.\mathbf{current}[\mathit{tid}(a')]$ = $s.\mathbf{current}[\mathit{tid}(a')] \cup \{\mathit{tid}(a)\}$; 
	}
}
\end{algorithm}
\vspace{-0.5cm}
\end{figure}

\begin{figure}[htb!]
\hrule
\begin{center}
\[
	\mathit{UpdateBack}(s, a)=
	\begin{cases}
		 \mathit{safeSet}(s), & \exists t \in s.\mathbf{current}[\mathit{tid}(a)] \setminus\mathit{safeSet}(s)\\
		 s.\mathbf{done}, & \exists T \subset s.\mathbf{done} : T =  \underset{t \in T}{\bigcup} s.\mathbf{current}[t]\\
		 \underset{t \in s.\mathbf{done}}{\bigcup} s.\mathbf{current}[t], & \textbf{otherwise}
	\end{cases}
\] 
\end{center}
\hrule
\noindent $s.\mathbf{current}[t]$: set of threads that
execute a transition dependent on $\mathit{next}(s,t)$ which appears after it in an execution. \\
\noindent $s.\mathbf{done}$: set of threads whose transitions have been fully explored from $s$.\\
\noindent $s.\mathbf{backtrack}$: when equal to $s.\mathbf{done}$, a source set for $s$.
\hrule
\vspace{-2mm}
\caption{Description of important components in Algorithm \ref{alg:eager}.}
\label{fig:alg2legend}
\vspace{-.3cm}
\end{figure}


Algorithm~\ref{alg:eager} builds on top of Algorithm \ref{alg:ample} by computing on-the-fly source sets to limit exploration of transitions from the \emph{irreducible} states.  
More precisely, reducible states are traversed according to the strategy of Algorithm \ref{alg:ample} (i.e., only one enabled invisible transition is followed) and  for irreducible states, source sets determine what transitions are followed. Since safe sets are also source sets, the overall algorithm remains sound if the new source sets are computed correctly.


Figure \ref{fig:alg2legend} provides a declarative description of the key components of Algorithm~\ref{alg:eager}. For a state $s$ in the current execution (stored on the stack), the $s.\mathbf{current}$ set may be updated every time a new visible transition is explored, and the $s.\mathbf{backtrack}$ set may be updated every time the exploration backtracks to $s$. The update of $s.\mathbf{backtrack}$ relies on the sets $s.\mathbf{current}$ computed while traversing successors of $s$.

When a new transition $(s,a,s')$ from a state $s$ is traversed, the active thread $\mathit{tid}(a)$ is added to the current set $s_{l}.\mathbf{current}[t]$,  where $s_l$ is the last state from which the current execution performs a transition that is dependent on $a$ such that $t\neq \mathit{tid}(a)$ is the thread of that transition. See line~\ref{line:curr} and the $\mathbf{UpdateCurr}$ function. When a transition is followed to a visited state $s$, the same update is done for \emph{every} transition that is reachable from $s$, as if these transitions are traversed again. See lines~\ref{line:start_revisit}-\ref{line:end_revisit} and note that the declarative definition of $A_s$ at line~\ref{line:start_revisit} corresponds to a traversal of all the executions starting from $s$. This may be time-consuming, and yet, such updates are unavoidable in stateful POR algorithms because the current execution reaching $s$ (stored on the stack) may belong to a different Mazurkiewicz trace compared to a previous execution reaching $s$ (whose sequence of transitions leading to $s$ was different).
When backtracking to a state $s$, the set $s.\mathbf{backtrack}$ is updated to take into account the transitions which are dependent on and occur after the last explored transition starting from $s$, called $\tau_s$. If $\tau_s$ is a transition of thread $t$, the threads performing those dependent transitions are stored in $s.\mathbf{current}[t]$. 
If there is a dependent transition $\tau$ performed by a thread $t'$ that is not in the safe set of $s$, then $s.\mathbf{backtrack}$ is updated conservatively to contain the safe set of $s$. This situation occurs when $\tau$ becomes enabled after executing some other thread $t''$ enabled in $s$, and observing an execution where $\tau_s$ occurs after $\tau$ requires first executing the transition of $t''$.
Otherwise, the algorithm checks to see if a subset $T$ of threads enabled in $s$ which have already been explored are sufficient to cover  $s.\mathbf{current}[t]$, and that $T$'s transitions in $s$ are independent from future transitions of threads not in $T$. 
In that case, $s.\mathbf{backtrack}$ is assigned with $s.\mathbf{done}$ and the exploration from $s$ is halted. The subset of threads $T$ defines a persistent set {\bf and} a source set for $s$. Since source sets are monotonic, $s.\mathbf{done}$ is a source set for $s$. If none of the previous conditions hold, then $s.\mathbf{current}[t]$ is simply added to $s.\mathbf{backtrack}$. This computation is defined by the macro $\mathit{UpdateBack}$ in Figure \ref{fig:alg2legend}.


\begin{figure}[t]
\begin{center}
\includegraphics[scale=0.6]{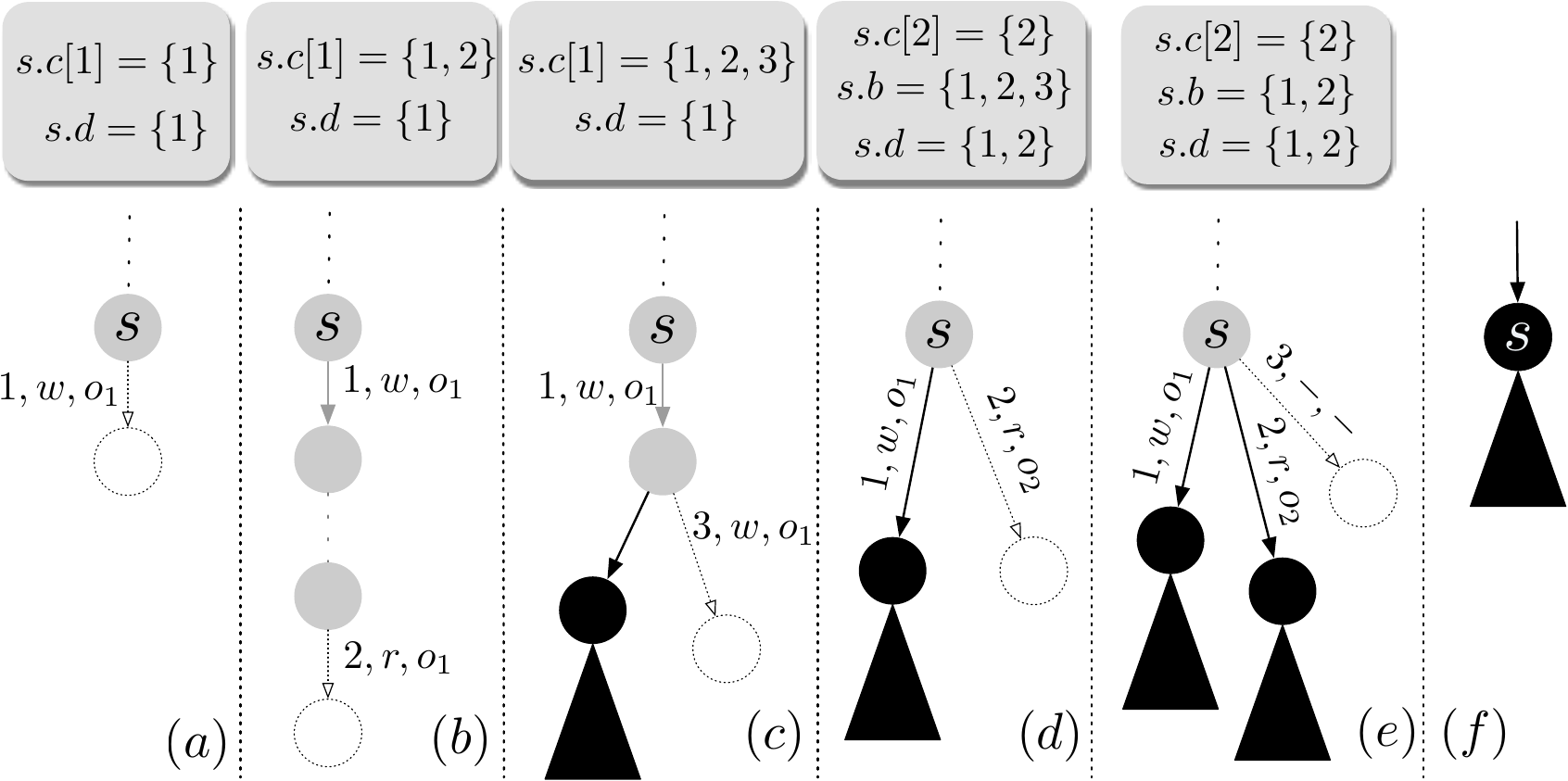}
\vspace{-4mm}
\caption{An example for Algorithm \ref{alg:eager}. Solid grey circles represent states stored on the stack, hollow dotted circles represent states on the top of the stack, and solid black circles represent states from which the exploration has been completed, i.e. their $\mathbf{backtrack}$ sets are equal to their $\mathbf{done}$ sets. Transitions follow the same pattern: dotted transitions are the latest to have been explored, solid grey ones are between states on the stack, and solid black ones are the ones taken in the past. Solid black triangles represent completed explorations starting from some state. We omit program counters from actions. $\mathbf{backtrack}$, $\mathbf{current}$, and $\mathbf{done}$ are abbreviated by the first letter.}
\label{Fig:flowStandardDiv}
\end{center}
\vspace{-0.3cm}
\end{figure}

We illustrate the algorithm using Figure~\ref{Fig:flowStandardDiv}. In $(a)$, $s$ is reached for the first time and the transition labeled by $(1,w,o_1)$ is selected first to be executed. This is a visible transition of thread 1 that writes to the shared object $o_1$.  
After this transition is taken, $s.\mathbf{current}[1]$ and $s.\mathbf{done}$ become $\{1\}$.  In $(b)$, from some state which is reached later, the transition labeled by action $(2,r,o_1)$ is selected to be followed next. Since this action is dependent on  $(1,w,o_1)$, thread $2$ is added to $s.\mathbf{current}[1]$. Then in $(c)$, a transition of thread $3$ with an action dependent on $(1,w,o_1)$ is taken, and $3$ is added to $s.\mathbf{current}[1]$.
After backtracking to $s$ in $(d)$, $s.\mathbf{backtrack}$ is updated by simply copying $s.\mathbf{current}[1]$. The next transition to be taken from $s$ belongs to thread 2 which is in $s.\mathbf{backtrack} \setminus s.\mathbf{done}$. This entails the  initialization of $s.\mathbf{current}[2]=\{2\}$ and the addition of $2$ to  $s.\mathbf{done}$.
In $(e)$, we backtrack to $s$ again and without having changed $s.\mathbf{current}[2]$. This means that $(2,r,o_2)$ is independent of any later action of another thread. Therefore, $\{2\}$ is a persistent set of $s$ and $s.\mathbf{done}=\{1,2\}$ a source set of $s$, and  $s.\mathbf{backtrack}$ is assigned with $s.\mathbf{done}$ to stop the exploration from $s$, as pictured in (f).

This example shows that Algorithm \ref{alg:eager} explores sets of transitions from a given state $s$ that may correspond to a source set which is \emph{not} a persistent set. The exploration  in Figure~\ref{Fig:flowStandardDiv} stops when $s.\mathbf{backtrack}=s.\mathbf{done}=\{1,2\}$, but the only persistent set that includes thread 1 is $\{1,2,3\}$.

\begin{theorem}\label{th:eager}
Given a program represented by an LTS $L$, Algorithm \ref{alg:eager} terminates with an LTS $L_r$ that is sound for $L$.
\end{theorem}
\label{sec:appendixProofEager}
\vspace{-3mm}
\begin{proof} Based on the soundness of source sets (see Section~\ref{ssec:por}), it is enough to show that for every state $s$ in $L_r$, 
\begin{align}
	& \parbox[t]{0.92\columnwidth}{\centering
	$s.\mathbf{backtrack}$ is a source set for $s$ in $L$ when it becomes equal to $s.\mathbf{done}$}
	\label{eq:proof_eager}
\end{align}
Due to the condition of \textbf{while} in line~\ref{line:backEqDone} of Algorithm~\ref{alg:eager}, equality of $s.\mathbf{backtrack}$ and $s.\mathbf{done}$ is the only condition for stopping an exploration from a state $s$. Therefore, if some successor state $s'$ of $s$ is already explored and the search is backtracked to $s$, then $s'.\mathbf{backtrack}=s'.\mathbf{done}$, because otherwise, \textbf{while} loop in line~\ref{line:backEqDone} wouldn't be terminated for $s'$. Since $s.\mathbf{done}$ keep tracks of threads whose enabled transitions from $s$ is already executed, the proof is reduced to showing that the following proposition holds:
\begin{align}
	& \parbox[t]{0.9\columnwidth}{\centering
	For any state $s$, $s.\mathbf{backtrack}$ is a source set when the exploration from $s$ is finished}
	\label{eq:proof_eager2}
\end{align}
If $s$ is a reducible state, then only one transition is explored from $s$ which is an invisible transition. The fact that the thread performing this invisible transition is a persistent set and hence a source set follows directly from definitions as every persistent set is also a source set. When $s.\mathbf{backtrack}=s.\mathbf{done}$ is different from $\mathit{safeSet}(s) = \mathit{enabled}(s)$ (which is trivially a source set), it must be the case that there exists $T \subset s.\mathbf{done}$ such that $T =  \underset{t \in T}{\bigcup} s.\mathbf{current}[t]$ due to the definition of $\mathit{UpdateBack}$ method. Now we show that $T$ is a persistent set for $s$ in $L$. Assume by contradiction that this is not the case, then due to the definition of persistent set, $L$ admits an execution $E$ starting from $s$ that contains only transitions of threads different from those in $T$ and at least one of these transitions $\tau$ of a thread $t'\not\in T$ is dependent on some transition $\mathit{next}(s,t)$ with $t\in T$. For every $t \in T$, the successor state $s'$ of $s$ reached by $\mathit{next}(s,t)$ must be in $L_r$. Due to deadlock freedom assumption, some transition that has the same transition label with $\tau$ must be enabled eventually in some successor state of $s'$. Let $E' \in L_r$ be that execution from $s$ which starts with $\mathit{next}(s,t)$ and contains such transition that shares the same label with $\tau$. 
Now we will move forward by showing that the following proposition is correct, which will be used in the rest of the proof:
\begin{align}
	& \parbox[t]{0.9\columnwidth}{\centering
	If $L_r$ admits an execution $E''$ from $s$ whose last transition that depends on and occurs before $\tau'$ is $\mathit{next}(s,t)$ where $act(\tau') = act(\tau)$ and $tid(\tau') = tid(\tau) = t'$, then $t' \in T$}
	\label{eq:proof_eager3}
\end{align} 
When $\tau'$ is executed from some successor state of $s$, $t'$ is added to $s.\mathbf{current}[t]$ (and eventually will be added to $T$ due to $\mathit{UpdateBack}$ method) by invoking the $\mathbf{UpdateCurr}$ function as $\mathit{next}(s,t)$ will be the transition in line~\ref{line:depExists} of Algorithm~\ref{alg:eager}. This contradicts the assumption of $t' \not\in T$ and therefore, it is enough to show that proposition below is correct for concluding the proof:
\begin{align}
	& \parbox[t]{0.92\columnwidth}{\centering
	If $L_r$ admits such an execution $E'$, then $L_r$ admits such an execution $E''$}	\label{eq:proof_eager4}
\end{align} 
To show that Proposition~\ref{eq:proof_eager4} holds, we proceed by induction on the order of $\mathit{next}(s,t)$ when we go backwards in $E'$. The base step is trivial since $E''$ can be $E'$ when $\mathit{next}(s,t)$ is the first transition. Assuming by induction that $\mathit{next}(s,t)$ is the $n$-th transition that is dependent on and occurs before $\tau'$ in $E'$ and $L_r$ admits such execution $E''$, we show that this also holds when $\mathit{next}(s,t)$ is ($n$+1)-th transition with the same properties. Let $s''$ be the state that is reached from $s$ by executing $E'_p$ which is the prefix of $E'$ until (not included) the last transition $\tau''$ that is dependent on and occurs before $\tau'$ and hence, $\tau''$ is enabled in $s''$. Using Proposition~\ref{eq:proof_eager3}, as $\mathit{tid}(\tau')$ must be in $T$ of $s''$, there must be another execution $E'''$ from $s''$ such that $\tau'''$ occurs before $\tau''$ where $act(\tau') = act(\tau''')$ and $tid(\tau') = tid(\tau''') = t'$ . Since in the execution starting from $s$ as $E'_p$ and continues as $E'''$, $\mathit{next}(s,t)$ is the $n$-th transition that is dependent on and occurs before $\tau'''$ (when we go backwards), proposition~\ref{eq:proof_eager4} is correct by using induction assumption. As mentioned, this contradicts the assumption of $t' \not\in T$ in proof of proposition~\ref{eq:proof_eager2} and thus, $T$ is a persistent set and a source set. By monotonicity of source sets, $s.\mathbf{backtrack}$ is also a source set. \qed
\end{proof}

\section{Lazy Source Set POR (DL-S-POR)}\label{sec:lazy}

%
%
%


\begin{figure}[ht!]
\begin{algorithm}[H]
\SetAlCapNameFnt{\scshape}
\SetAlgoVlined
\caption{Lazy Source Set POR ({\lazy}) } 
\label{alg:lazy}
\SetKwInput{init}{Initialize}{}{}
\SetKwProg{search}{Explore}{()}{end}
\SetKwProg{safeSet}{safeSet}{(s)}{end}
\SetKwProg{isComplete}{IsComplete}{(s)}{end}

\init{$Stack \gets \emptyset$; $Stack.\mathbf{push}(s_I)$; $L_r \gets \emptyset$;}

\search{}{
	$s \gets Stack.\mathbf{top}$;\\
	$s.\mathbf{backtrack}  \gets \emptyset$; $s.\mathbf{done} \gets \emptyset$; $s.\mathbf{current} \gets \emptyset$;\\
	\While {true \label{line:backDone}} {
		\uIf{$\exists t_1\in s.\mathbf{backtrack} \setminus s.\mathbf{done}$ \label{line:lazy:selectStart}}{
			$t\gets t_1$
		}
		\uElse{
			choose $t\in \mathit{safeSet}(s)\setminus s.\mathbf{done}$\label{line:lazy:selectEnd}
		}
		$(s,a,s')=\mathit{next}(s,t)$;\\
		$Stack.\mathbf{push}(s')$;\\
		\If {$\mathit{notVisited}(s')$ \label{line:lazy:not_visited}} {
			$\mathbf{Explore}()$;
		}
		\If {$\mathbf{IsComplete}(s)$ \label{line:lazy:isComplete}} { 
			$Stack.\mathbf{pop}()$;\\
			\Return \label{line:return}
		}
		$Stack.\mathbf{pop}()$;\\
	}
}

\isComplete{}{
	\ForAll {$(s, a, s') \in \{s' \in L_r : t=\mathit{tid}(a)\not\in s.\mathbf{done} \}$\label{line:lazy:curr_condition}} {
		$s.\mathbf{done}$ = $s.\mathbf{done}\cup \{t\}$;\label{line:done}\\
		$T \gets \mathit{safeSet}(s)$;\\
		\If {$s.\mathbf{done} = T$ $\lor$ $(\forall t' \in T: \mathit{isVisited}(\mathit{succ}(s,t')) \lor t' \in s.\mathbf{done})$ \label{line:lazy:opt}} {
			add transitions $(s,a,s')$ to $L_r$;\label{line:lazy:add_trans}\\
			$s.\mathbf{done} \gets T$;\\
			$s.\mathbf{backtrack} \gets s.\mathbf{done}$;\\
			\Return true;
      		}
		$s.\mathbf{current}[t] \gets \{t\}$;\label{line:lazy:start_current}\\
		$A_{s'} \gets \{a': \mbox{$a'$ occurs in an execution from $E(L_r,s')$}\}$;\label{line:lazy:traversal_current}\\ 
		$s.\mathbf{current}[t]$ = $s.\mathbf{current}[t]\cup \{\mathit{tid}(a'): a'\in A_{s'} \mbox{ and }a \nsim a'\})$;\label{line:lazy:end_current}\\
		$s.\mathbf{backtrack} \gets \mathit{UpdateBack}(s, a)$;\label{line:lazy:back}\\
		\If {$s.\mathbf{backtrack} = s.\mathbf{done}$ \label{line:completeRet}} {
			\Return true;
		}
	}
	\Return false;
}
\end{algorithm}
\vspace{-1.1cm}
\end{figure}

Algorithm~\ref{alg:eager} tracks dependencies between transitions in an eager manner, i.e., every new transition leads to updates of $\mathbf{current}$ sets. In this section, we present a lazy variation that computes such dependencies only when the exploration backtracks to a state. The incentive is to compute such dependencies only when needed to decide if the exploration from a given state should continue or not. 
 Also, this enables several optimizations when traversing the state space to compute such dependencies that are not possible in the eager version. 

Algorithm~\ref{alg:lazy} presents our POR algorithm based on a lazy computation of source sets. 
%
Rather than updating the $\mathbf{current}$ sets on-the-fly for states on the stack, this algorithm re-traverses part of the state space each time it backtracks to a state $s$ in order to update just $\mathbf{current}$ sets of $s$. This is done in the function $\mathbf{IsComplete}$.
As a result, $s.\mathbf{done}$ is populated with a new thread $t$ just before computing dependencies with $t$'s transition in $s$ and not after executing that transition in the style of Algorithm~\ref{alg:eager} (see line~\ref{line:done}). For every transition  $\tau$ of a thread $t$ from $s$ that has been followed since the last time the algorithm backtracks to $s$ (i.e., $t$ is not already in $s.\mathbf{done}$ -- see line~\ref{line:lazy:curr_condition}), the algorithm updates $s.\mathbf{current}[t]$ to include all threads $t'$ that later execute a transition that is dependent on $\tau$ (see lines~\ref{line:lazy:start_current}-\ref{line:lazy:end_current}). Subsequently, the $s.\mathbf{backtrack}$ set is updated exactly in the style of  Algorithm~\ref{alg:eager} (see line~\ref{line:lazy:back}). If $s.\mathbf{backtrack}$ becomes equal to $s.\mathbf{done}$ then $\mathbf{IsComplete}$ returns $\mathit{true}$ and the exploration from $s$ stops.


\begin{figure}[t]
\centering
\includegraphics[scale=0.6]{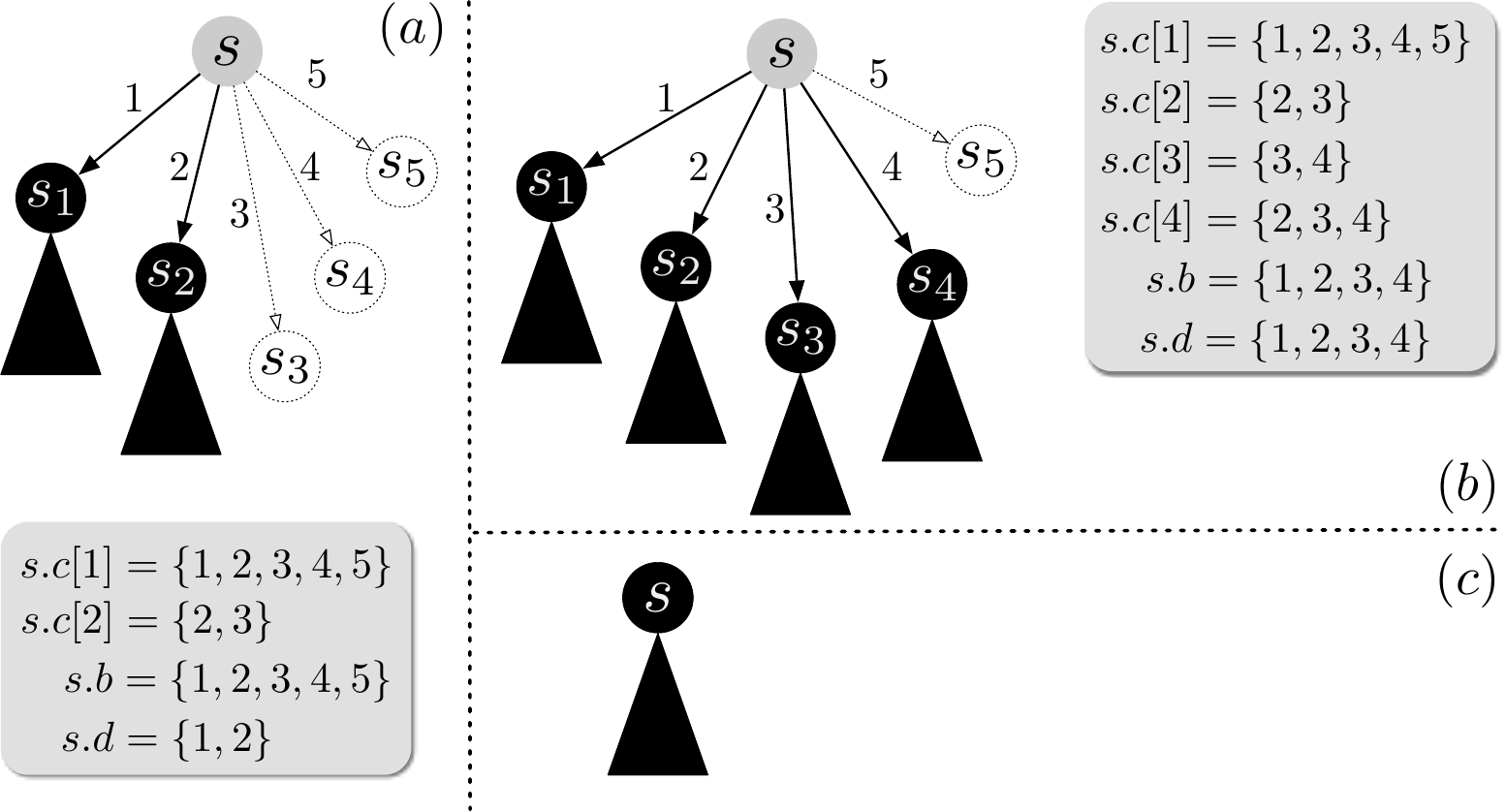}
\vspace{-5mm}
\caption{An example exploration of Lazy Source Set POR. We use the same conventions as in Figure~\ref{Fig:flowStandardDiv}.}
\label{Fig:flowBUComplete}
\vspace{-3mm}
\end{figure}

We explain how Algorithm~\ref{alg:lazy} works by an example.  Figure~\ref{Fig:flowBUComplete}(a) illustrates a scenario in which the exploration backtracks to a state $s$ for the second time. After the first backtrack to $s$, the state space starting from the successor $s_1$ (resulted from following a transition of thread 1) was re-traversed in order to compute $s.\mathbf{current}[1]$. We assume that $s.\mathbf{current}[1]$ is changed to $\{1, 2, 3, 4, 5\}$ due to the dependent transitions encountered during this traversal. The set $s.\mathbf{backtrack}$ is set to $s.\mathbf{current}[1]$ as the latter contains all the enabled transitions. The exploration continues with a transition from $s$ of thread $2$ which is possible because thread 2 is in $s.\mathbf{backtrack} \setminus s.\mathbf{done}$. After backtracking to $s$ for the second time, the re-traversal of the state space starting in $s_2$ leads to $s.\mathbf{current}[2]=\{2, 3\}$. The set $s.\mathbf{backtrack}$ remains the same after this computation. Then, in Figure \ref{Fig:flowBUComplete}(b), when backtracking to $s$ for the fourth time, we assume that $s.\mathbf{current}[3]=\{3, 4\}$ and $s.\mathbf{current}[4]=\{2, 3, 4\}$. Since transitions of threads 2, 3, and 4 starting in $s$ are independent of transitions of other threads that occur later, we can conclude that $\{2,3,4\}$ is a persistent set and $\{1,2,3,4\}$ is a source set, and update $s.\mathbf{backtrack}$ to $s.\mathbf{done}$. Therefore, the exploration from $s$ stops, as pictured in Figure~\ref{Fig:flowBUComplete}(c). The set of transitions explored from $s$ corresponds to a source set which is not a persistent set. The only persistent set that includes thread 1 is $\{1,2,3,4,5\}$.



In both {\eager} and {\lazy}, we optimize re-traversals after some state $s$ (computing $A_s$ at line \ref{line:start_revisit} and line \ref{line:lazy:traversal_current}, respectively) by not traversing all the executions after $s$ but just traversing each transition after $s$ only once. {\lazy} is also amenable to other optimizations that are not possible or difficult to implement for {\eager}. These optimizations for {\lazy} either prevent some re-traversals inside the $\mathbf{IsComplete}$ method (see the if block at line 21) or provide early exit conditions for them. All these optimizations are explained in detail in Appendix~\ref{sec:appendixRetraverse}.

The soundness of Algorithm~\ref{alg:lazy}, stated in the following theorem, is also based on proving that every state is expanded according to a source set. As in Theorem~\ref{th:eager}, it can be shown that $s.\mathbf{backtrack}$ is a source set for $s$ when it becomes equal to $s.\mathbf{done}$. When backtracking to a state, the $\mathbf{current}$ sets satisfy the same specification as in the eager version. 

\vspace{-.5mm}
\begin{theorem}\label{th:lazy}
Given a program represented by an LTS $L$, Algorithm \ref{alg:lazy} terminates with an LTS $L_r$ that is sound for $L$.
\end{theorem}
\label{sec:appendixProofLazy}
\vspace{-3.5mm}
\begin{proof} Similar to the proof of Theorem~\ref{th:eager}, it is enough to show that Proposition~\ref{eq:proof_eager} holds. To end an exploration from a state $s$, \textbf{while} loop in line~\ref{line:backDone} of Algorithm~\ref{alg:lazy} must be terminated. For this, $\mathbf{return}$ statement in line~\ref{line:return} must be reached and therefore, $\mathbf{IsComplete}$ in line~\ref{line:lazy:isComplete} should return true. First, we show that $\mathbf{IsComplete}$ method eventually returns true. Due to to method $\mathit{UpdateBack}$, we know that $s.\mathbf{backtrack}$ can not contain a thread $t \not\in \mathit{safeSet(s)}$. Hence, all the transitions of $s$ that can be executed are only from threads in $\mathit{safeSet(s)}$ because of lines~\ref{line:lazy:selectStart}--\ref{line:lazy:selectEnd}. Each time a transition of $s$ is executed and then the search backtracks to $s$, $\mathbf{IsComplete}(s)$ is initiated. By the \textbf{for} loop in line~\ref{line:lazy:curr_condition}, every transition from $s$ that is executed is considered and by line~\ref{line:done}, we know that all these transitions will be added to $s.\mathbf{done}$. Thus, $s.\mathbf{done}$ eventually becomes equal to $\mathit{safeSet(s)}$ which satisfies the condition in line~\ref{line:lazy:opt} and as a result, $\mathbf{IsComplete}$ method returns true. 

For $\mathbf{IsComplete}$ method to return true, either condition in line~\ref{line:lazy:opt} or line~\ref{line:completeRet} should be satisfied, where in both conditions $s.\mathbf{backtrack}$ must be equal to $s.\mathbf{done}$ before the return statement. That's why, equality of $s.\mathbf{done}$ and $s.\mathbf{back}$-\\$\mathbf{track}$ is the only condition for stopping an exploration from a state $s$. Similar to Algorithm~\ref{alg:eager}, since $s.\mathbf{done}$ keep tracks of threads whose enabled transitions from $s$ is already executed, the proof is deduced to showing that Proposition~\ref{eq:proof_eager2} holds for Algorithm~\ref{alg:lazy} as well. The part between Proposition~\ref{eq:proof_eager2} and Proposition~\ref{eq:proof_eager3} in the proof of Theorem~\ref{th:eager} applies totally the same and we show that $T$ is a persistent set for $s$ in $L$ using the fact that $L_r$ admits such an execution $E'$ as it is concluded in the same proof. Assume by contradiction that $T$ is not a persistent set for $s$. 
	But as a result of backtracking to $s$ after executing $E'$ and invoking $\mathbf{IsComplete}$ method, $t'$ will be added to $s.\mathbf{current}[t]$ (and eventually to $T$ due to $\mathit{UpdateBack}$ method) in line~\ref{line:lazy:end_current} of Algorithm~\ref{alg:lazy} since the transition label of $\tau'$ is an element of $A_{s'}$ ($t' \not\in T$ and $act(\tau) \nsim act(\tau')$) in line~\ref{line:lazy:traversal_current}. Since it contradicts the assumption, $T$ (and also $s.\mathbf{backtrack}$ by monotonicity of source sets) is a persistent set and a source set. \qed
	\end{proof}
\vspace{-6mm}
\subsection{Experimental Evaluation}
\label{sec:exp}

We evaluate an implementation of the three algorithms {\ample}, {\eager}, and {\lazy}, presented in Section~\ref{sec:eager} and Section~\ref{sec:lazy}, in the context of the Java Pathfinder (JPF) model checker. As benchmark, we use bounded-size clients of Java concurrent data structures.

\vspace{-5mm}
\subsubsection{Implementation}
\label{sssec:implementationPor}

We implement our algorithms as an extension of the \texttt{DFSHeur}-\\\texttt{istic} class in JPF. To identify (in)visible actions (for computing safe sets), the only manual input is a list of class names that constitute the implementation of the concurrent data structure. The (in)visible transitions are automatically inferred from these class names and Java synchronization-related native methods used to implement compare-and-swap (CAS) for instance, which are all known. Every action reading or writing a field of an object in one of these classes, or which corresponds to a native method call are marked as visible (JPF makes it possible to parse the Bytecode instructions executed in a transition and determine the read/written object fields). Calls to the \texttt{lock} and \texttt{unlock} methods of a lock object are both considered as writes to the lock object, and therefore, visible. Any other action is considered as invisible.
The dependency relation between visible actions is defined as usual, i.e., two actions that access the same object field, one of them being a write, are considered dependent. The way we define (in)visible actions is sound because the clients we consider do not contain additional computation. They simply call methods of the data structure (from different threads), the verification goal being related to combinations of return values observed in their executions.

\vspace{-5mm}
\subsubsection{Benchmarks}
\label{sssec:benchmarksPor}

Our benchmark consists of bounded-size clients of 7 concurrent data structures from JDK8 or Synchrobench~\cite{DBLP:conf/ppopp/Gramoli15}: two set implementations based on \emph{coarse-grain} and \emph{fine-grain} locking, respectively (RWLockCoarseGrainedList-\\IntSet and OptimisticListSortedSetWaitFreeContains), a set implementation ba-\\sed on a binary search tree and CAS, 
a wrapper on top of java.util.concurrent.Co-\\ncurrentLinkedQueue, 
java.util.concurrent.ConcurrentHashMap and a wrapper on top of it,
 and a hash map implementation based on coarse-grain locking.
Since these implementations update shared memory using compare-and-swap or guarded by locks, they are data-race free and the restriction to sequential consistency is sound.

To evaluate our algorithms, we sampled 75 clients of these data structures where each client calls add and remove methods from 3 threads. Each thread contains up to 5 calls. 
We varied the contention on shared objects using less or more distinct inputs for add and remove methods.

We also use a number of buggy variations of the lock-based sets, RWLockCoarseGrainedListIntSet and OptimisticListSortedSetWaitFreeContains. We us-\\ed Violat to generate client programs of these variations that admit consistency violations. Violat generates these client programs in three steps. First, Violat enumerates arbitrary test programs of a given data structure based on other inputs such as number of threads, maximum number of programs and so on. Next, it computes expected (\textsc{adt}-admitted return-value) outcomes for each test program by computing and then recording the outcomes of all possible sequential executions. Finally, it runs the threads of each test program in parallel (using a stress testing tool or JPF), checks if the results are as expected, and reports the test programs that violates linearizability which is witnessed by observing an unexpected outcome.

To introduce bugs in the selected data structures before inputting them to Violat, we modify the placement of locks dynamically under certain conditions in certain methods (e.g., when the set contains a specific element). These conditions make it possible to control the difficulty of a bug.  
We consider four different classes of clients based on the number of invocations to methods that lead to bugs: (1) all of the invocations, (2) half of the invocations, (3) just a single invocation and (4) none of the invocations. We sampled 310 clients of these buggy variations with 3 threads and up to 4 calls per thread using Violat.
  

\vspace{-5mm}
\subsubsection{Results}
\label{sssec:resultsPor}


We use {\ample}, {\lazy}, {\eager} to denote the three algorithms presented in this paper. For the same algorithms, we use JPF, DL-JPF, DE-JPF to represent the standard setup of JPF, and variations of the {\lazy} and {\eager} when the safe set of a state $s$ contains all the enabled threads in $s$ ($\mathit{safeSet}(s) = \mathit{enabled}(s)$). The latter are used to evaluate the performance of the eager and lazy approaches while disabling the benefit of the static S-POR method. We compare implementations of {\ample}, {\lazy} and {\eager} between them, with JPF, DL-JPF and DE-JPF, with their stateless variations, and with a stateful variation of the optimal source set algorithm in~\cite{DBLP:journals/jacm/AbdullaAJS17} (called O-DPOR). For a fair comparison, we implement O-DPOR on top of {\ample} without wakeup trees as their operations are quite expensive. The experiments were run on a 2,3 GHz Dual-Core Intel Core i5 processor with 8GB of RAM. We consider a timeout of 30 minutes.

\begin{figure}[ht]
\vspace{-6mm}
  \centering
  \includegraphics[width=\columnwidth]{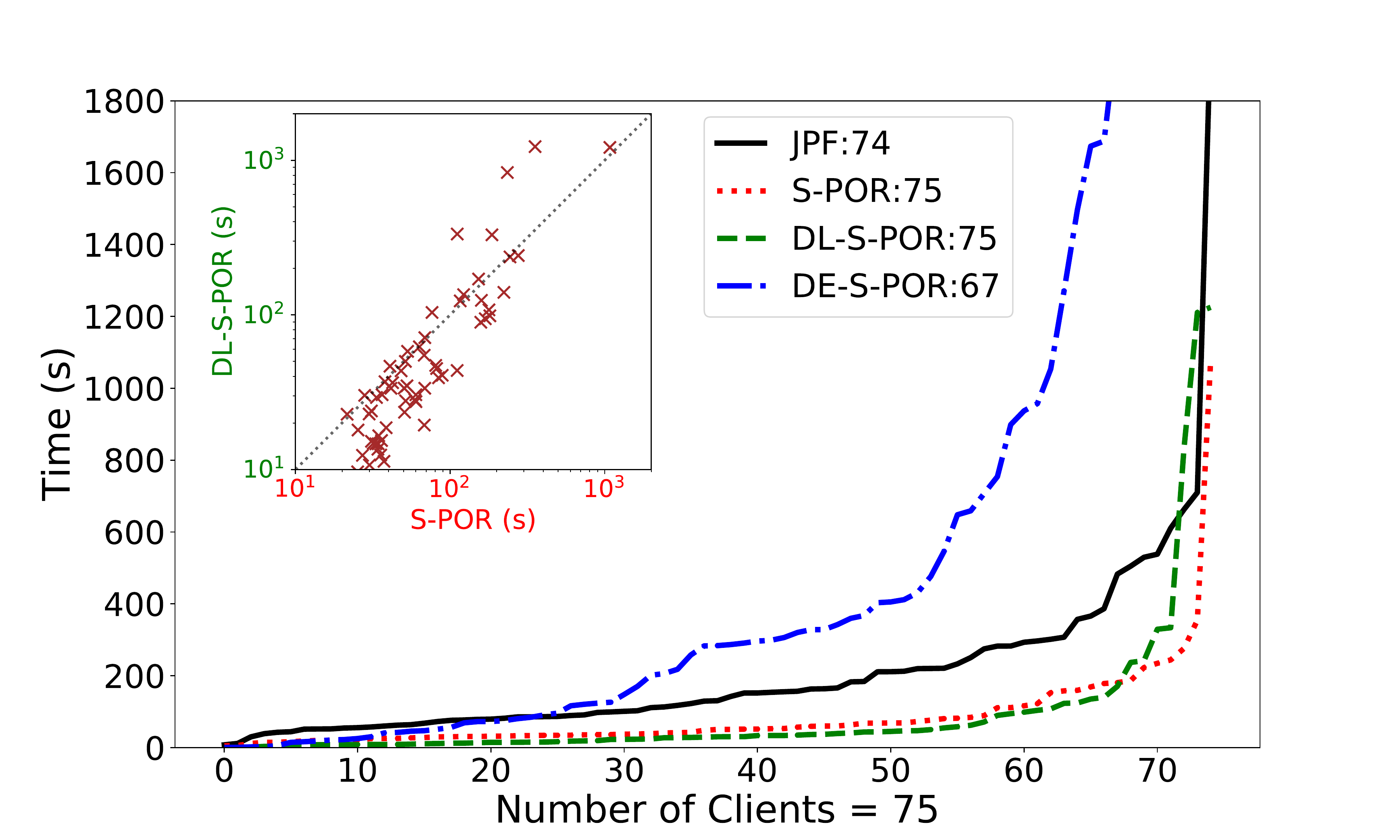}
  \vspace{-7mm}
  \caption{Quantile plot of running times for {\ample}, {\lazy}, {\eager} and JPF (for each algorithm, clients are ordered w.r.t. time in ascending order). The top left part shows a scatter plot for comparing {\ample} and {\lazy}.}
  \label{fig:compareAll}
  \vspace{-3mm}
\end{figure}

\begin{figure}[h]
  \centering
  \includegraphics[width=\columnwidth]{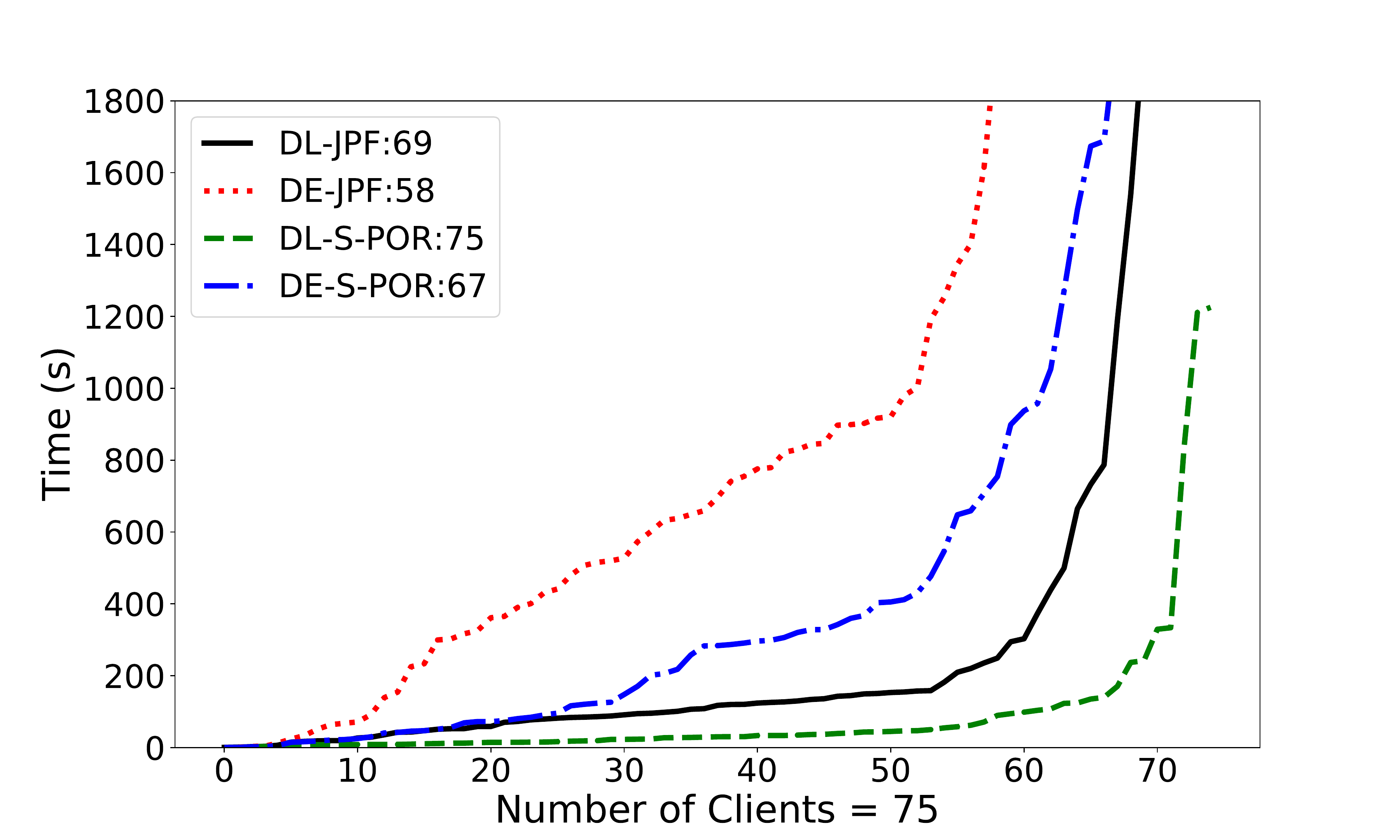}
  \vspace{-7mm}
  \caption{Quantile plot of running times for DL-JPF, DE-JPF, {\lazy} and {\eager} (for each algorithm, clients are ordered w.r.t. time in ascending order as in Figure~\ref{fig:compareAll}).}
  \label{fig:compareAll2}
\end{figure}

\noindent
\textbf{Execution time comparison.}
Figure~\ref{fig:compareAll} and Figure~\ref{fig:compareAll2} present a comparison in terms of execution time between different sets of algorithms. In Figure~\ref{fig:compareAll}, we compare JPF, {\ample}, {\lazy} and {\eager} to observe the advantages of using our algorithms against the standard setup of JPF. In Figure~\ref{fig:compareAll2}, we compare {\lazy}, {\eager}, DL-JPF and DE-JPF for investigating the gain by applying static filtering using {\ample} as a baseline in dynamic algorithms. To ease the interpretation of the results, for each algorithm, we order clients according to execution time in ascending order. The numbers in the legend represent the number of clients on which a given algorithm terminates before the timeout. 
We omit O-DPOR because it times out for a large part of the benchmark, i.e., 39 out of the 46 clients on which it was run (our implementation of the algorithm in~\cite{DBLP:journals/jacm/AbdullaAJS17}  does not support programs using locks which makes it inapplicable to the rest of the clients). This optimal algorithm manipulates happens-before constraints between steps in an execution, which results in a large overhead compared to our simpler tracking of pairwise dependencies. 
We also omit stateless variations of our algorithms since none of them finished before the timeout for any client. Note that stateless versions are obtained by disabling the state matching\footnote{JPF uses hashing for state matching which is theoretically imperfect and can lead to incomplete results on rare occasions} in JPF, which also disables storing the full reachability graph.

Results based on Figure~\ref{fig:compareAll} show that the lazy source set computation in {\lazy} gives a significant speedup w.r.t. {\eager} (and intuitively O-DPOR) while outperforming JPF. While {\ample} processes few more clients faster w.r.t. {\lazy}, 
the scatter plot on the top-left of  Figure~\ref{fig:compareAll} shows that it is mostly in favor to {\lazy} when clients are observed individually (this plot is given in logarithmic scale). {\lazy} performs better than {\ample} if there is a high potential for reduction, i.e., the ratio between the number of states explored by {\lazy} over {\ample} is smaller, and otherwise, {\ample} is the best. This supports the hypothesis that if the potential for reduction is high enough then a carefully customized dynamic computation  of source sets has a significant impact on performance. {\lazy} gives an average speedup (average of speedups for each client) of 2.6 compared to {\ample}. Overall picture suggests using a portfolio model checker where {\ample} and {\lazy} are run in parallel.

Similar to Figure~\ref{fig:compareAll}, Figure~\ref{fig:compareAll2} illustrates a comparison in terms of time between {\lazy}, {\eager}, DL-JPF and DE-JPF. It shows that our algorithms outperforms their variations that are directly built on top of JPF ({\lazy} against DL-JPF and {\eager} against DE-JPF). It also highlights the fact that the lazy approach is still better than the eager one even when the lazy approach is not based on {\ample}.

\begin{figure}[ht]
\vspace{-7mm}
  \centering
  \includegraphics[width=0.98\columnwidth]{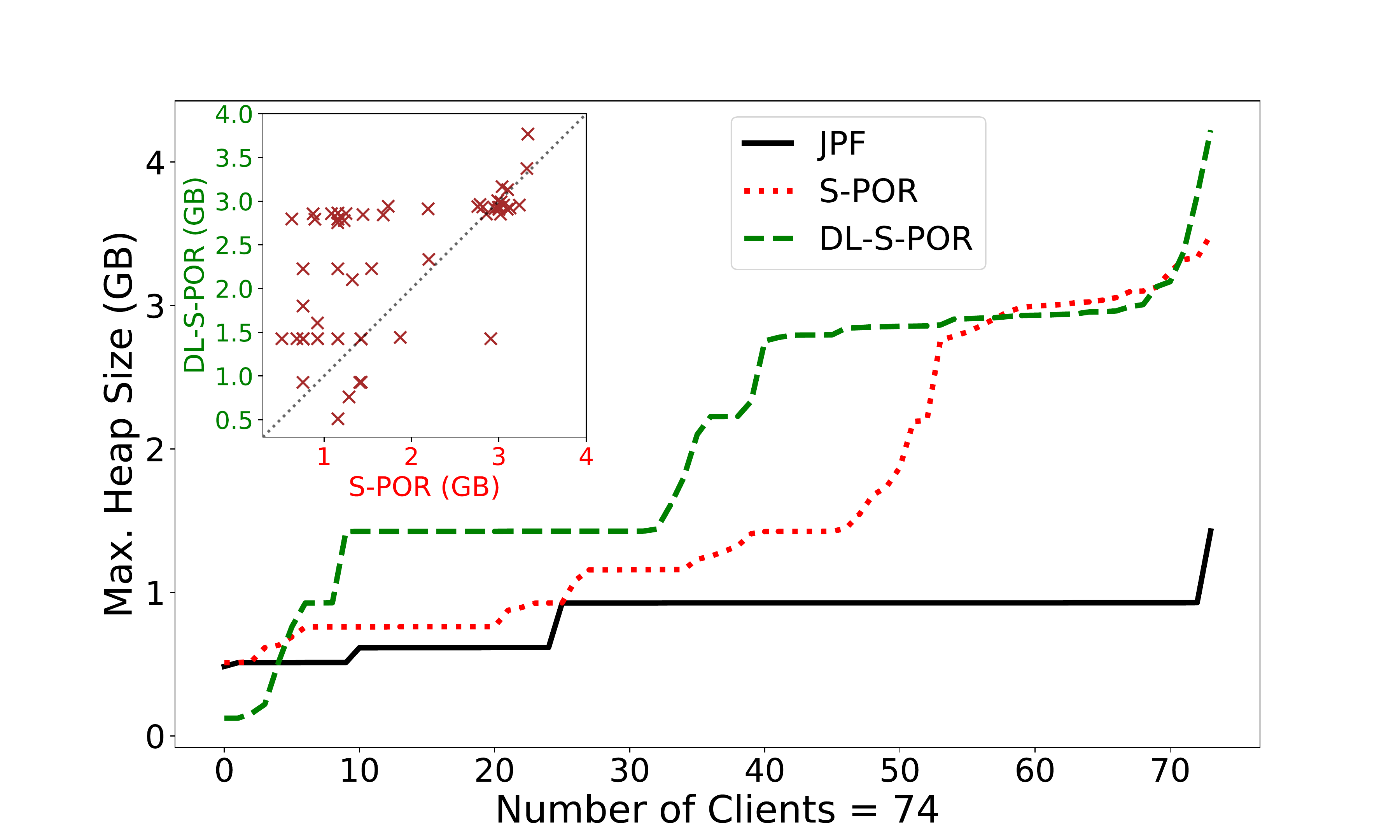}
  \vspace{-4mm}
  \caption{Quantile plot of memory consumption for {\ample}, {\lazy}, and JPF (for each algorithm, clients are ordered w.r.t. memory in ascending order). The top left part shows a scatter plot for comparing {\ample} and {\lazy}.}
  \label{fig:memory}
\end{figure}

\begin{figure}[ht]
  \centering
  \includegraphics[width=0.98\columnwidth]{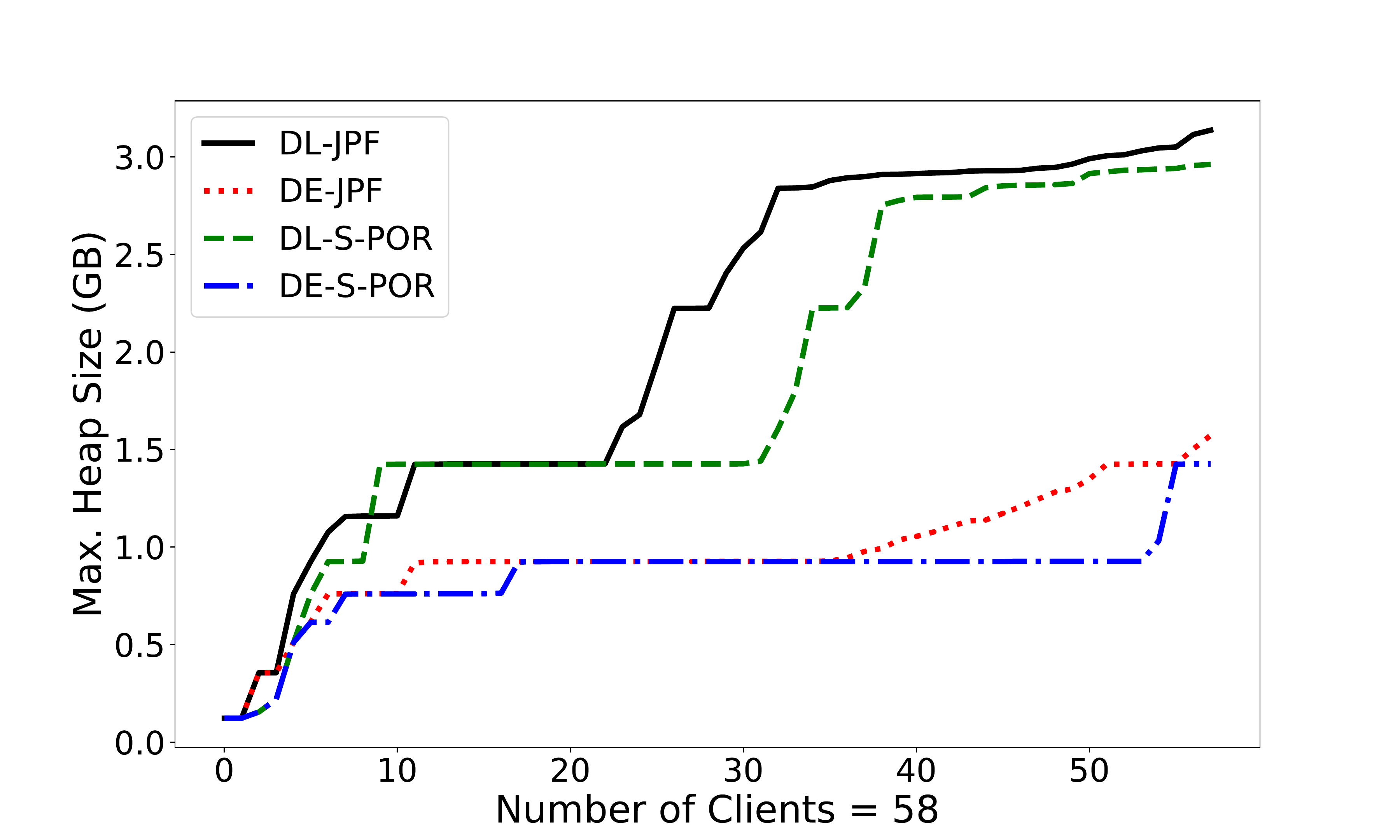}
  \vspace{-4mm}
  \caption{Quantile plot of memory consumption for DL-JPF, DE-JPF, {\lazy} and {\eager} (for each algorithm, clients are ordered w.r.t. memory in ascending order as in Figure~\ref{fig:memory}).}
  \label{fig:memory2}
\end{figure}

\noindent
\textbf{Memory consumption comparison.}
Figure~\ref{fig:memory} presents a comparison in terms of memory consumption between {\ample} and {\lazy}, the most efficient algorithms according to Figure~\ref{fig:compareAll} and Figure~\ref{fig:compareAll2}, 
against the standard setup of JPF. We compared the maximum heap sizes using 74 clients that terminate before timeout for all algorithms. In all the experiments, the highest allocated heap size is 4.2GB. {\ample} and {\lazy} consume more memory than JPF because they have to store the transition labels which are used to reduce the explored state space. This overhead is unavoidable for any form of dynamic partial order reduction. However, this memory consumption overhead is counterbalanced by significant speedups in terms of time. There is some memory overhead also due to storing the sets of transition labels manipulated by the algorithms, e.g., $s.\mathbf{current}$. But since these sets are maintained only for irreducible states and they are deleted for a state $s$ when $s.\mathbf{done}$ equals $s.\mathbf{backtrack}$, their effects are not significant as storing transition labels. 

For 32\% of the clients, {\ample} and {\lazy} consume at most twice the memory consumed by JPF. For these clients, the average memory overhead is 1.00 for {\ample} and 1.34 for {\lazy} while the average speedup against JPF is 2.54 and 6.67, respectively.
For 50\% of the clients, {\ample} and {\lazy} consume in between 2 and 4 times the memory used by JPF. The average memory overhead for these clients is 2.20 for {\ample} and 2.63 for {\lazy} while the average speedup is 2.86 and 7.81, respectively. 
For the rest of the clients, the memory overhead is at most 7.79 and in average 4.11 for {\ample} and 5.39 for {\lazy} while the average speedup 3.28 and 5.31, respectively.
 
The top-left part of Figure~\ref{fig:memory} shows a pair-wise comparison of allocated maximum heap sizes in {\ample} and {\lazy}. These algorithms are incomparable in general. After investigating the clients individually, the results confirm that {\lazy} consumes less memory than {\ample} when there is a high potential for reduction. The memory consumed for computing source sets is compensated by the reduction in the state space.

When we take a look at to Figure~\ref{fig:memory2}, it illustrates a comparison between {\lazy}, {\eager}, DL-JPF and DE-JPF as in Figure~\ref{fig:compareAll}, but in terms of memory consumption. As in Figure~\ref{fig:memory}, we compared the maximum heap sizes of clients that finished before it timed out for all algorithms, which are 58 of them. It demonstrates that our algorithms are slightly better than their variations that are directly built on top of JPF. This memory overhead is mainly because of the additional transition labels that are not removed by the static filter.  The overhead is also due to storing the sets of transition labels manipulated by the algorithms for all of the states rather than just for the irreducible ones but as mentioned previously, this overhead is negligible. This figure also shows that tracking dependencies with an eager approach does not increase the heap size as much as the lazy approach, although they explore the same state space. 
This difference in the memory overhead can not be explained by the memory that is used for storing the sets of transition labels or LTSs as they are all the same for both algorithms and they are kept in the same data structures. We suspect that this overhead can be due to low-level, internal details of JPF or due to the garbage collection process which might not keep up.

\begin{figure}[t]
\vspace{-5mm}
  \centering
  \begin{minipage}[b]{0.49\columnwidth}
    \includegraphics[width=\columnwidth]{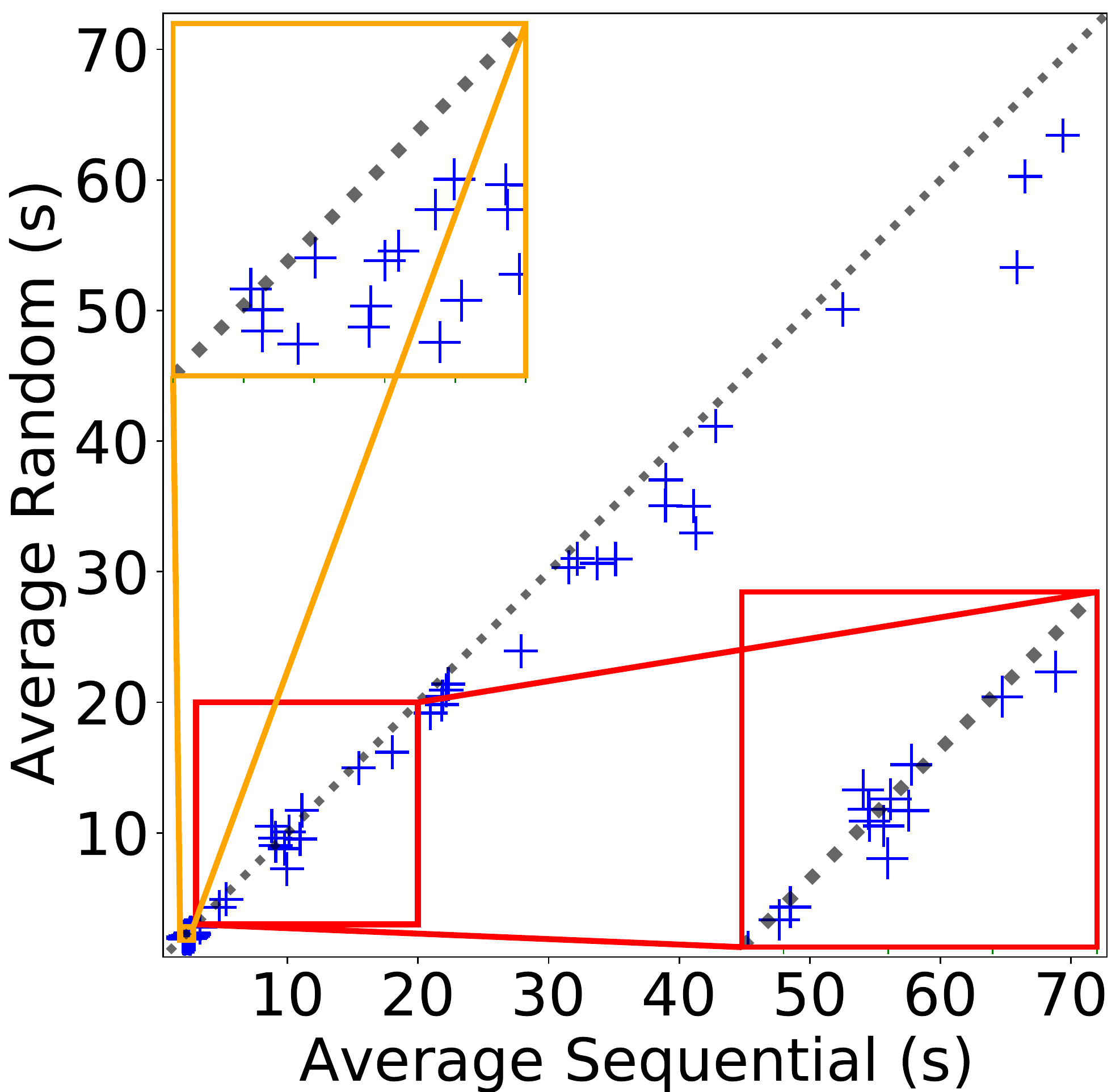}
  \end{minipage}
  \hfill
  \begin{minipage}[b]{0.49\columnwidth}
    \includegraphics[width=\columnwidth]{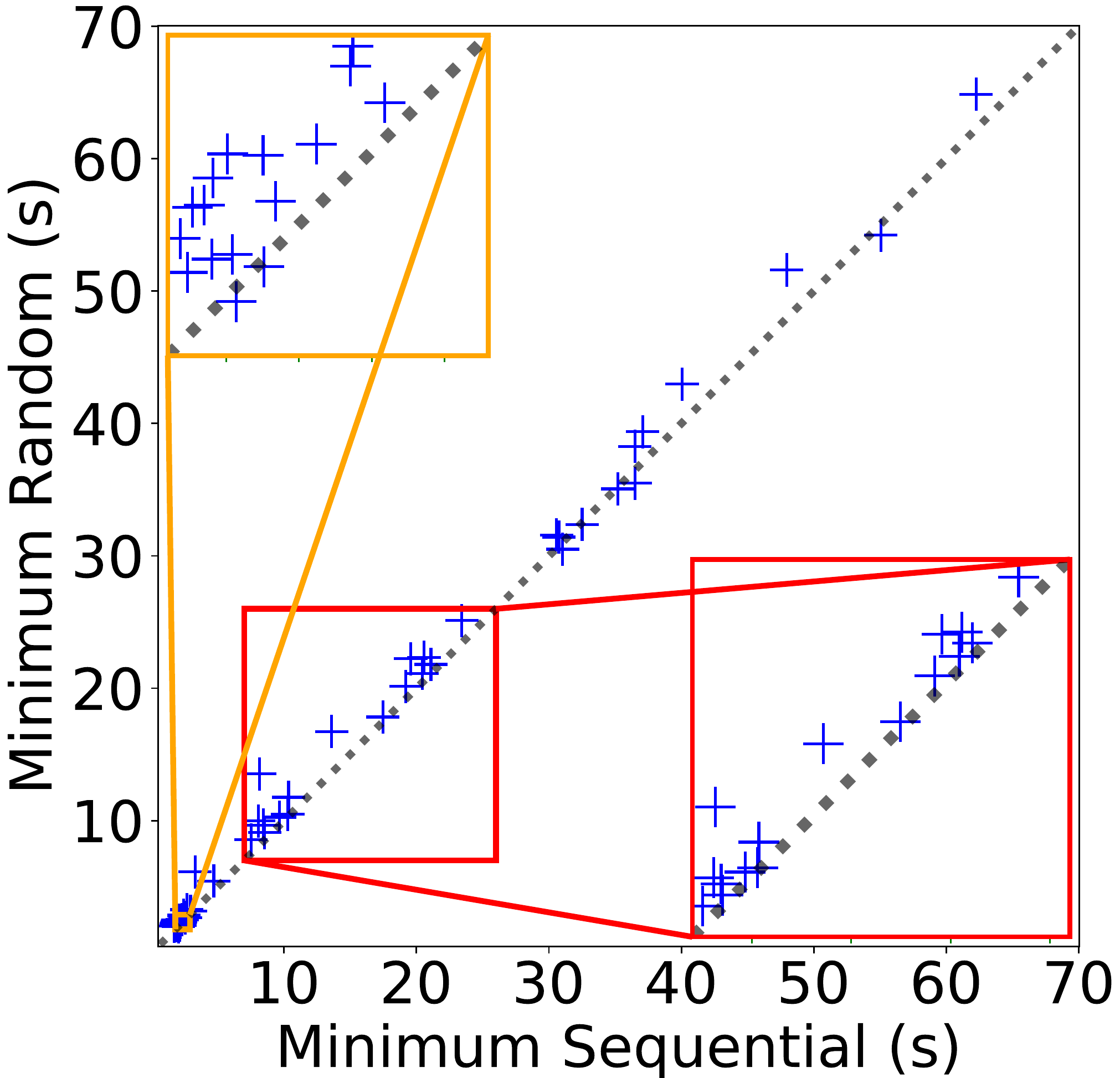}
  \end{minipage}
   \vspace{-3mm}
  \caption{Time comparison between the sequential and random strategies when {\lazy} enumerates all states, using clients with a single buggy invocation.}
   \label{Fig:AllLess-main}
\end{figure}

\noindent
\textbf{Transition enumeration.}
The performance of dynamic POR algorithms is generally affected by the order in which transitions starting in a certain state are enumerated. This order influences the size of the computed persistent/source sets. This order is also important when enumerating states only until the first error is detected. We evaluate two strategies for defining this order, called {\emph{sequential}} and {\emph{random}}. For both of these strategies, the algorithm first selects a transition that leads to an already visited state, if one exists. We made this choice because it leads to better performance (this is adopted by the standard setup of JPF as well). In the sequential strategy, if there is no such transition, then the algorithm picks a transition by respecting a pre-defined order between the thread ids. In random, the next transition is selected uniformly at random. 

We ran {\ample} and {\lazy}, the best algorithms as shown above, with all 6 permutations of the 3 threads for sequential and 3 different seeds for random. For each strategy, we report the average and minimum time over different instances. 

\begin{figure}[t]
  \centering
  \begin{minipage}[b]{0.49\columnwidth}
    \includegraphics[width=\columnwidth]{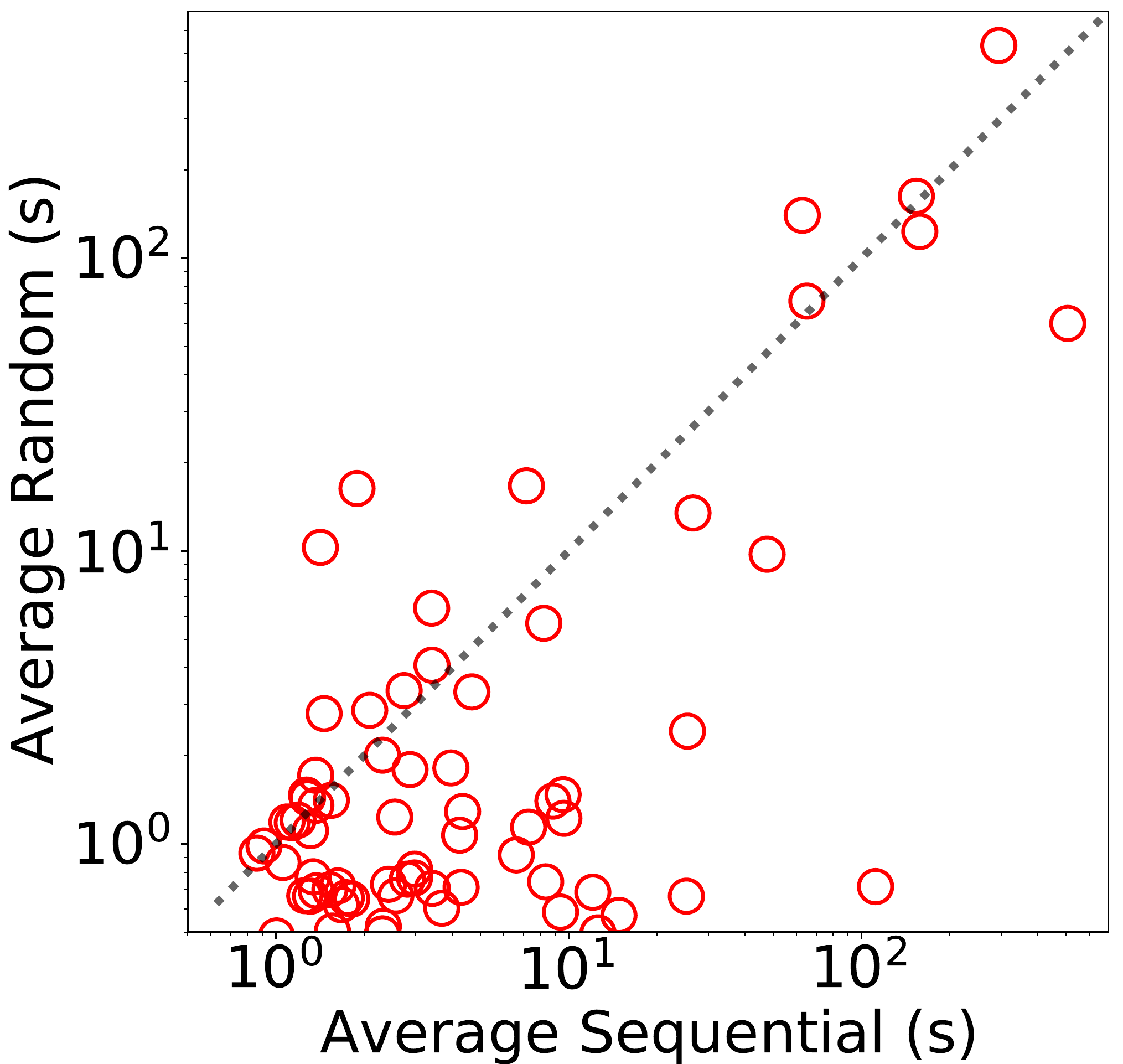}
  \end{minipage}
  \hfill
  \begin{minipage}[b]{0.49\columnwidth}
    \includegraphics[width=\columnwidth]{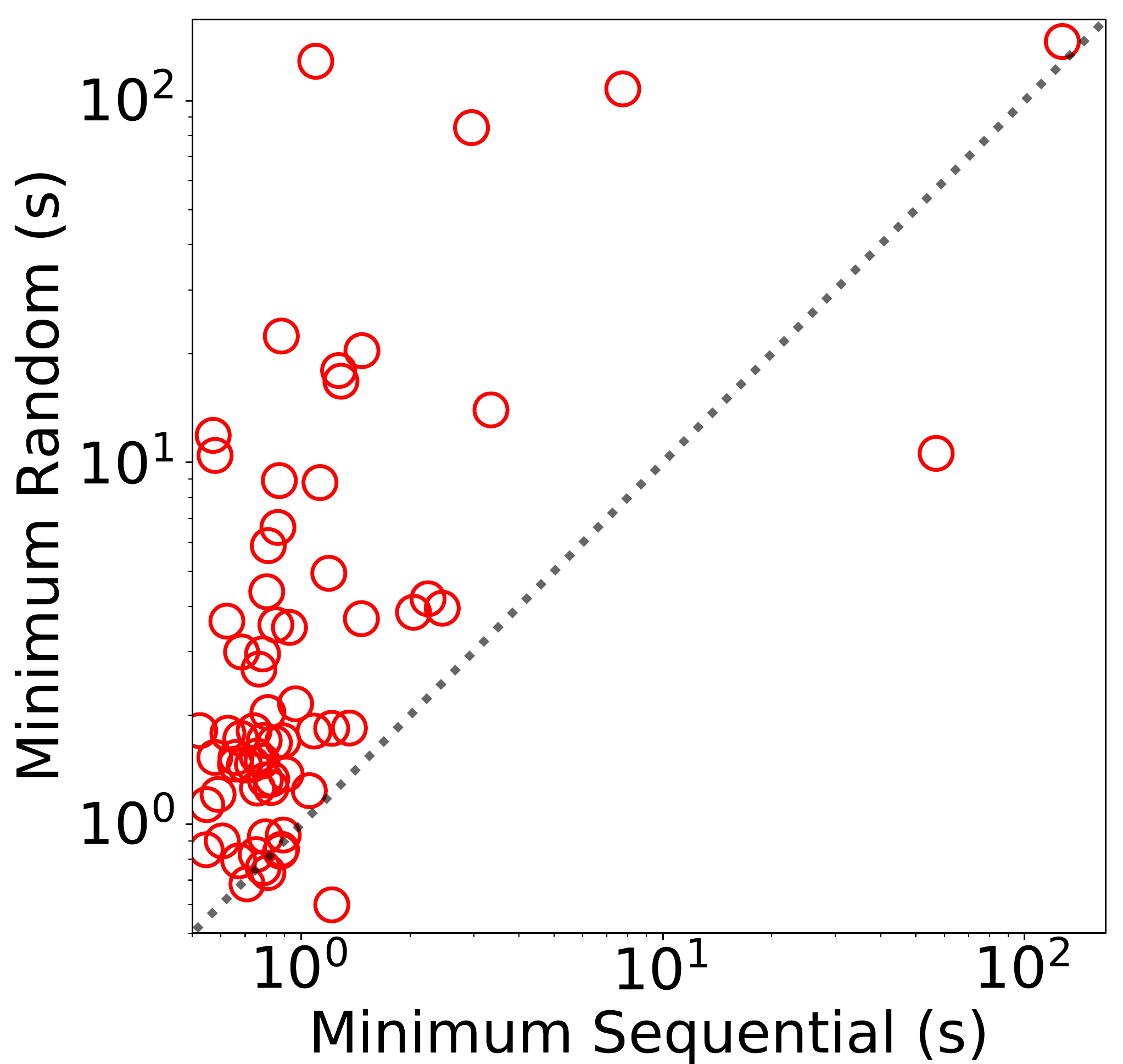}
  \end{minipage}
   \vspace{-3mm}
   \caption{Time comparison (log scale) between the sequential and random strategies when {\lazy} enumerates states only until the first error, using clients in which all the invocations are buggy.}
  \label{Fig:FirstLazyMore-main}
\end{figure}

We report on the impact of these enumeration strategies for  {\lazy} when computing \emph{all} reachable states (up to POR) in Figure~\ref{Fig:AllLess-main} and \emph{only until the first error} in Figure~\ref{Fig:FirstLazyMore-main}. For {\ample}, the enumeration strategy is not important for the first case (since there is no dynamic computation of persistent/source sets) and it has a similar effect as for {\lazy} in the second case. These figures consider clients of buggy libraries where a single or all method invocations are buggy. The rest of the cases are presented in Appendix~\ref{sec:appendixOtherResults} and are similar.

The results show that the random strategy performs better in average, shown on the left of Fig.~\ref{Fig:AllLess-main}--\ref{Fig:FirstLazyMore-main}, but worse w.r.t. minima, shown on the right  of Fig.~\ref{Fig:AllLess-main}--\ref{Fig:FirstLazyMore-main}. 

The differences are more significant when enumerating states only until the first error. Note that Fig.~\ref{Fig:FirstLazyMore-main} reporting on this case is given in logarithmic scale. Thus, the sequential strategy should be preferred when using a portfolio model checker, i.e., parallel runs for each permutation of thread ids, and otherwise, the random strategy is better. This follows also from the average standard deviation being 28 seconds for random and 60 seconds for sequential, where the means are 17 and 20 seconds, resp. Note that, there is no significant impact observed from changing the algorithms or the number of buggy method invocations in the clients.
\vspace{-1mm}
\section{Related Work}\label{sec:related}


Over the years various different techniques have been introduced to deal with the state explosion problem in model checking. For concurrent programs specifically, depth bounding \cite{DBLP:conf/popl/Godefroid97}, delay bounding \cite{DBLP:conf/popl/EmmiQR11}, context bounding (bounding the number of context switches) \cite{DBLP:conf/tacas/QadeerR05}, preemption bounding \cite{DBLP:conf/pldi/MusuvathiQ07} and phase bounding \cite{DBLP:conf/tacas/BouajjaniE12} bring tractability to the model checking problem and have been shown to be effective for bug finding. These techniques are all {\em incomplete}, in the sense that lack of bugs does not guarantee the correctness of the system.

POR techniques reduce the search space by not exploring multiple executions from the same equivalence class, and are {\em complete}. Early techniques like {\em ample sets} \cite{DBLP:conf/popl/ClarkeES83,DBLP:conf/forte/HolzmannP94} and {\em stubborn sets} \cite{DBLP:journals/fmsd/GodefroidW93,DBLP:conf/dimacs/Godefroid90} were based on static analysis. {\em Sleep sets} \cite{DBLP:conf/dimacs/Godefroid90} were the first to  guarantee optimality (one execution from each equivalence class) \cite{DBLP:journals/fmsd/GodefroidHP95} by keeping track of information from the history of the exploration. However, they only prune transitions and cannot eliminate any state when used alone. Persistent sets \cite{DBLP:conf/cav/GodefroidP93,DBLP:journals/dc/KatzP92} generalized stubborn and ample sets and enabled development of dynamic POR (DPOR) methods.

In \cite{DBLP:conf/popl/FlanaganG05}, an efficient stateful algorithm is proposed for computing persistent sets dynamically by considering  currently explored parts of the state space. This algorithm needed large memory for keeping discovered states and the happens-before relation. The algorithm is improved in~\cite{DBLP:conf/spin/YangCGK08} with a more efficient state representation, and in \cite{DBLP:conf/icfem/YiWY06} with a summary-based representation of the happens before.


In \cite{DBLP:conf/fase/LauterburgKMA10,DBLP:conf/forte/TasharofiKLLMA12}, stateless dynamic POR techniques were introduced. {\em Source sets} \cite{DBLP:journals/jacm/AbdullaAJS17} were introduced in the context of dynamic POR techniques such that the state space can be reduced up to the limit that is theoretically possible. They are generalizations of persistent sets and their relation with persistent sets are investigated in \cite{DBLP:conf/birthday/AbdullaAJS17}. Our {\eager} and {\lazy} algorithms are relying on source sets but operate in the context of stateful model checking. The technique from \cite{DBLP:conf/atva/NeeleWBP16} is similar to our {\ample} algorithm for the GPU setting, but their choice of invisible actions is different than ours.





While we focus on shared-memory programs running on top of a sequential consistency memory model, POR techniques have been also investigated in the context of weak memory models such as TSO or C11, e.g.,~\cite{DBLP:conf/cav/Kokologiannakis21,DBLP:conf/asplos/Kokologiannakis20,DBLP:journals/acta/AbdullaAAJLS17,DBLP:journals/pacmpl/AbdullaAJN18}.
\vspace{-1mm}
\section{Conclusions}
\vspace{-1mm}


We proposed two algorithms for stateful model checking based on POR which build on the recently proposed source sets. Our algorithms focus on overall performance instead of theoretical optimality. Their evaluation in the context of JPF shows that they outperform a theoretically optimal algorithm~\cite{DBLP:journals/jacm/AbdullaAJS17}, and a simple static POR algorithm when there is a big enough potential for reducing the state space. This suggests that an effective model checker would have to run {\ample} and {\lazy} in parallel, and depending on the amount of parallelism resources available, with different instances of the sequential or random strategies for enumerating transitions starting in a certain state.


Reductions based on Mazurkiewicz trace equivalence~\cite{DBLP:conf/ac/Mazurkiewicz86} has also been used in proof simplification for concurrent verification \cite{DBLP:journals/pacmpl/FarzanV20} hypersafety verification \cite{DBLP:conf/cav/FarzanV19}. A relevant problem of interest is the automatic inference of inductive invariants \cite{DBLP:conf/pldi/MiltnerPMW20,DBLP:conf/popl/0001NMR16,DBLP:journals/fmsd/SharmaA16} that prove the correctness of concurrent libraries. The sound LTS's that are computed in this paper for the verification of individual instances, each provide a data point in what the inductive invariant for a most general client may look like. The key observation is that the inductive invariant for a most general client under some reduction scheme may be substantially simpler than an inductive invariant for all executions of the most general client. An interesting direction of future work is to investigate if the results of a sequence of individual client tests can be generalized to the discovery of a complete invariant (and hence a complete proof) under an appropriate reduction scheme. {\lazy} provides an efficient way to produce data for a data-driven inference algorithm.
%
%
%
\bibliographystyle{splncs04}
\bibliography{dblp,misc}
\newpage
\appendix
\section{Optimizations}\label{sec:appendixRetraverse}
The re-traversals used to track dependencies in $\mathbf{current}$ sets are time consuming, but several optimizations can be applied to {\eager} and {\lazy}. First optimization is the only one that is applied to both of these algorithms, which is traversing each transition only once rather than traversing all the executions after some state $s$ for calculating $A_s$ at line \ref{line:start_revisit} and line \ref{line:lazy:traversal_current}, respectively. This approach is much cheaper in general due to the high ratio of visited states and can be achieved easily by disabling (re-)visiting an already visited state during the same re-traversal. For guaranteeing soundness, we need to update $\mathbf{UpdateCurr}$ function for re-traversals in {\eager} to find not just the last transition (as it is stated at line~\ref{line:depExists}) but all the transitions that their actions are dependent on $a$ for conservatively updating $\mathbf{current}$ sets for all such transitions (at line 22). Such an update is needed since some of the different orderings between transitions can be missed during these optimized re-traversals. On the other hand, {\lazy} does not need such an update for guaranteeing soundness after this optimization as it already updates $\mathbf{current}$ sets conservatively (see line~\ref{line:lazy:end_current}). Rest of the optimizations in this section are only applied to {\lazy}.

\begin{figure}[h]
\centering
\includegraphics[scale=0.6]{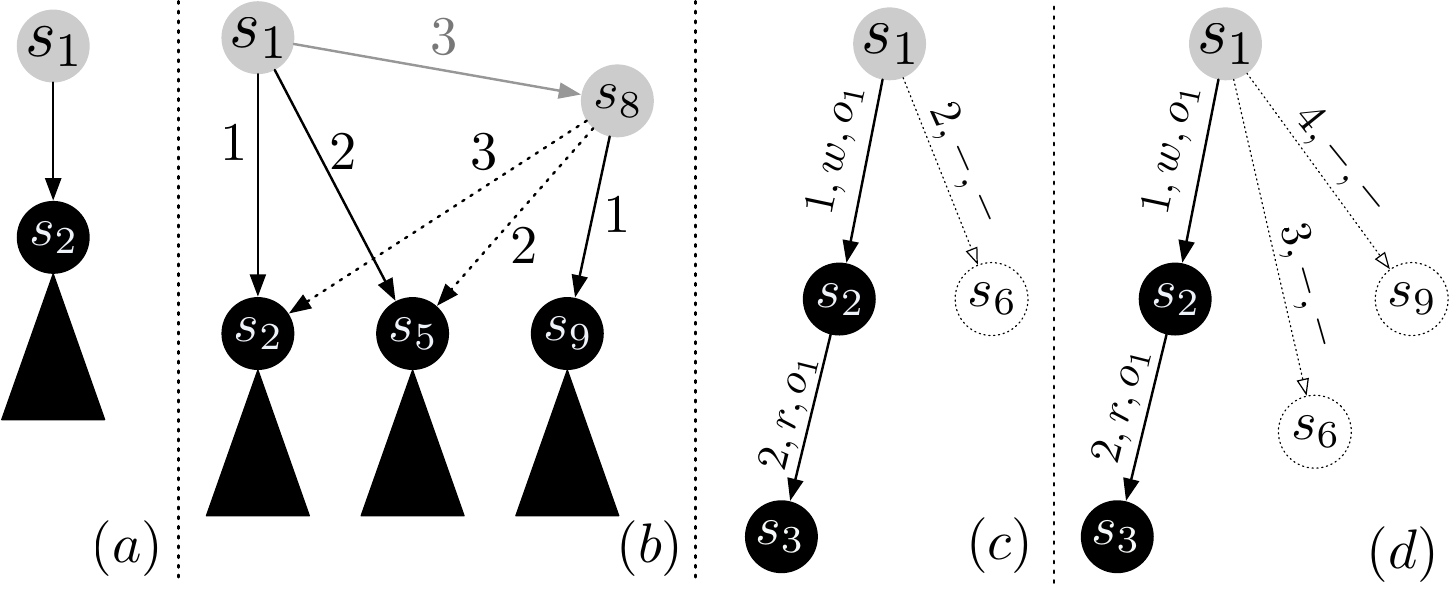}
\vspace{-2mm}
\caption{Optimizations to avoid re-traversing the state space. We use the same conventions as in Figure~\ref{Fig:flowStandardDiv}.}
\label{Fig:noDFS}
\vspace{-5mm}
\end{figure}

The simplest optimization for {\lazy} is not performing a traversal from the last successor of $s$, called $s'$, when backtracking to $s$. Then, $s.\mathbf{done}$ becomes equal to $safeSet(s)$ after adding the thread $t$ leading to $s'$ (see the first disjunct at line~\ref{line:lazy:opt}). This implies that no re-traversal is initiated for a state $s$ that is reducible or it is irreducible but has only one enabled transition; for an example see Figure~\ref{Fig:noDFS}(a).

Second, a traversal from a successor of $s$ does not have to be performed if after adding $t$ to $s.\mathbf{done}$, all the threads enabled in $s$ and not already in $s.\mathbf{done}$ lead to already visited states (see the second disjunct at line~\ref{line:lazy:opt}). The successors of those transitions may have not been pushed to the stack so they are added to the current LTS $L_r$ (see line~\ref{line:lazy:add_trans}). This optimization also relies on initiating re-traversals of the state space only when backtracking from a new state (see the condition at line~\ref{line:lazy:not_visited}). Stopping the traversal is sound because there are no new states to explore from $s$ and the exploration from $s$ is already complete. A scenario where this optimization is enabled is shown in Figure~\ref{Fig:noDFS}(b). When backtracking to $s_8$ for the first time (from $s_9$), all the other threads enabled in $s_8$ lead to an already visited state. The re-traversal of the state space starting in $s_9$ is stopped since the exploration from $s_8$ is already complete (before returning, the transitions of threads 2 and 3 from $s_8$ are added to the current LTS).

While these optimizations for {\lazy} stop re-traversals altogether, several optimizations concern stopping a traversal early before reaching every state. In a concrete implementation, the declarative definitions at lines~\ref{line:lazy:traversal_current}-\ref{line:lazy:end_current} translate to a (DFS) traversal of all the transitions starting in $s'$ and populating $s.\mathbf{current}[t]$ as new dependent transitions are found. This traversal can be stopped as soon as $s.\mathbf{current}[t]$ becomes ``complete'', i.e., it stores all threads in $\mathit{safeSet}(s)$. An example is shown in Figure~\ref{Fig:noDFS}(c) where the traversal starting in $s_2$ can stop immediately after the first transition since only threads 1 and 2 are enabled in $s_1$. Similarly, the traversal can be stopped immediately as $s.\mathbf{current}[t]$ contains a thread which is not enabled in $s$. In this case, $s.\mathbf{backtrack}$ will anyway be updated conservatively to include all the threads in $\mathit{safeSet}(s)$. An example is shown in Figure~\ref{Fig:noDFS}(d).

Only the first optimization is adapted in the eager version and it requires an update for soundness. Rest of these optimizations are not even applicable in the eager version, or if they are, they are much harder to apply. Here, we take advantage of having to compute dependencies using forward traversals that explore transitions starting in a certain state as opposed to a backward traversal of states in the stack. 


\section{More Experimental Results}
\label{sec:appendixOtherResults}
\begin{figure}[!tbp]
  \centering
  \begin{minipage}[b]{0.49\columnwidth}
    \includegraphics[width=\columnwidth]{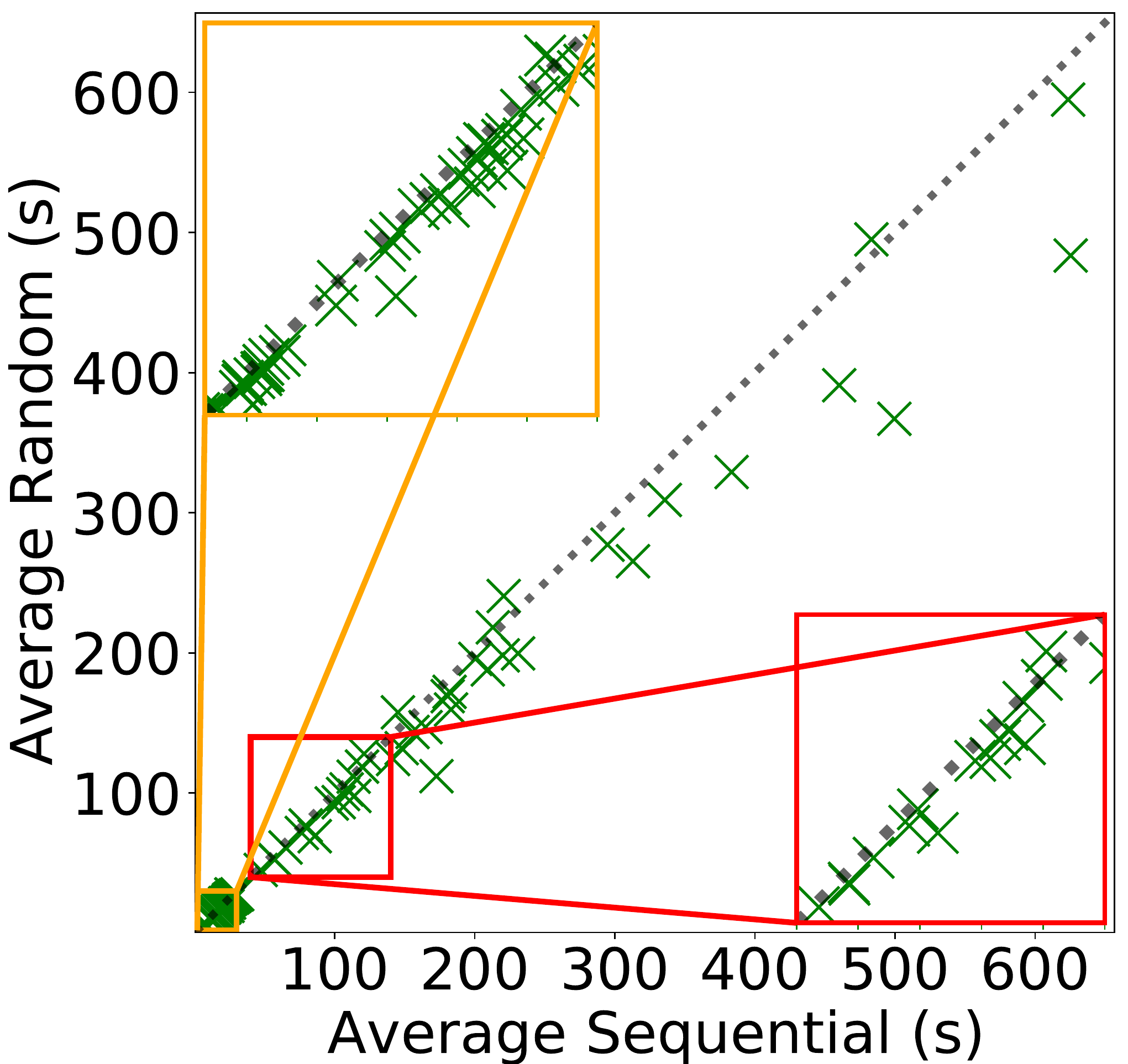}
  \end{minipage}
  \hfill
  \begin{minipage}[b]{0.49\columnwidth}
    \includegraphics[width=\columnwidth]{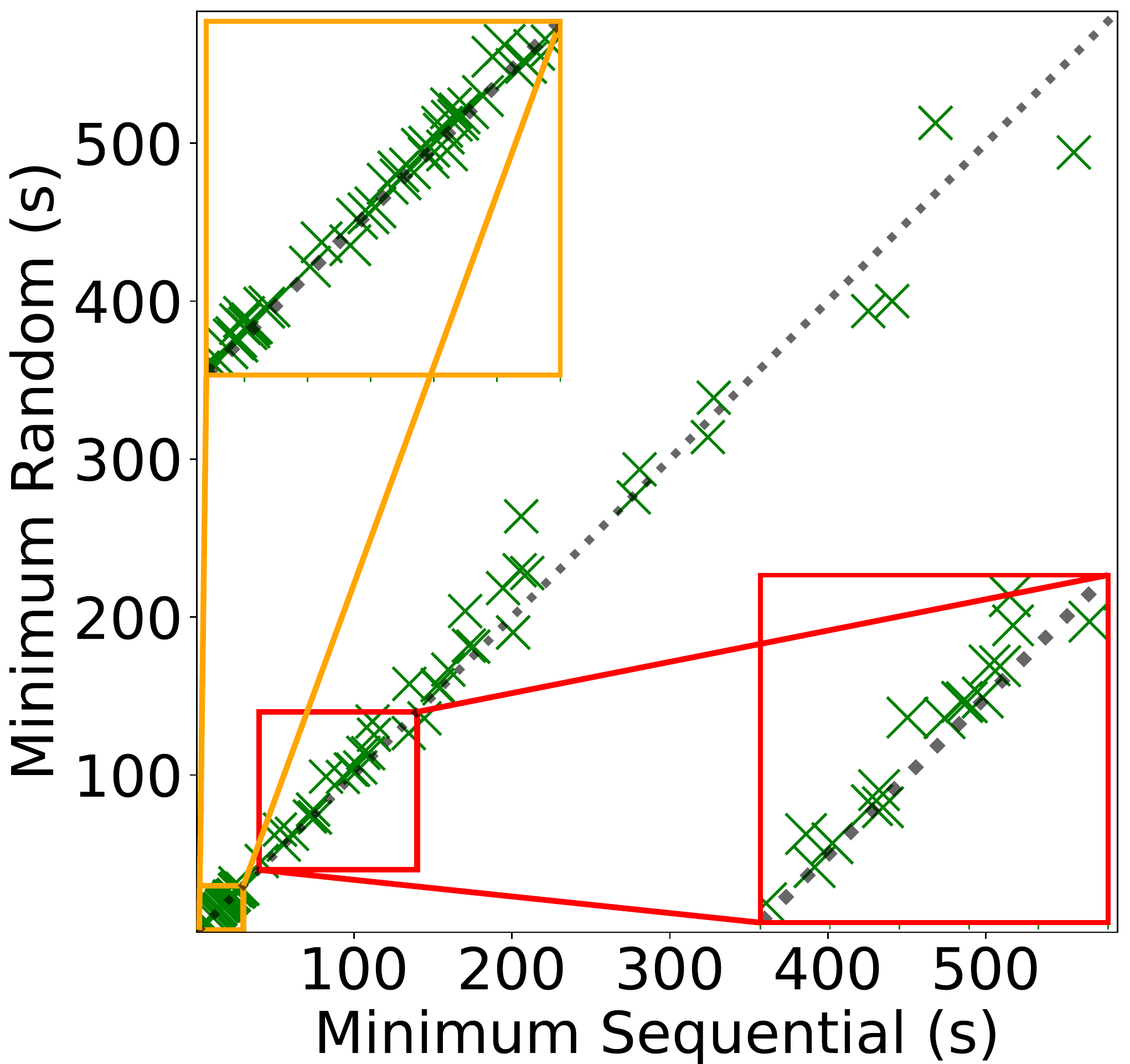}
  \end{minipage}
  \caption{{\textbf{ALL STATES $\backslash$ DL-S-POR $\backslash$ BUG: NO INVOCATION}} \newline Time comparison between sequential and random strategies when {\lazy} computes all states, using clients in which none of the invocations are buggy.}
  \label{Fig:AllCorrect}
\end{figure}

\begin{figure}[!tbp]
  \centering
  \begin{minipage}[b]{0.49\columnwidth}
    \includegraphics[width=\columnwidth]{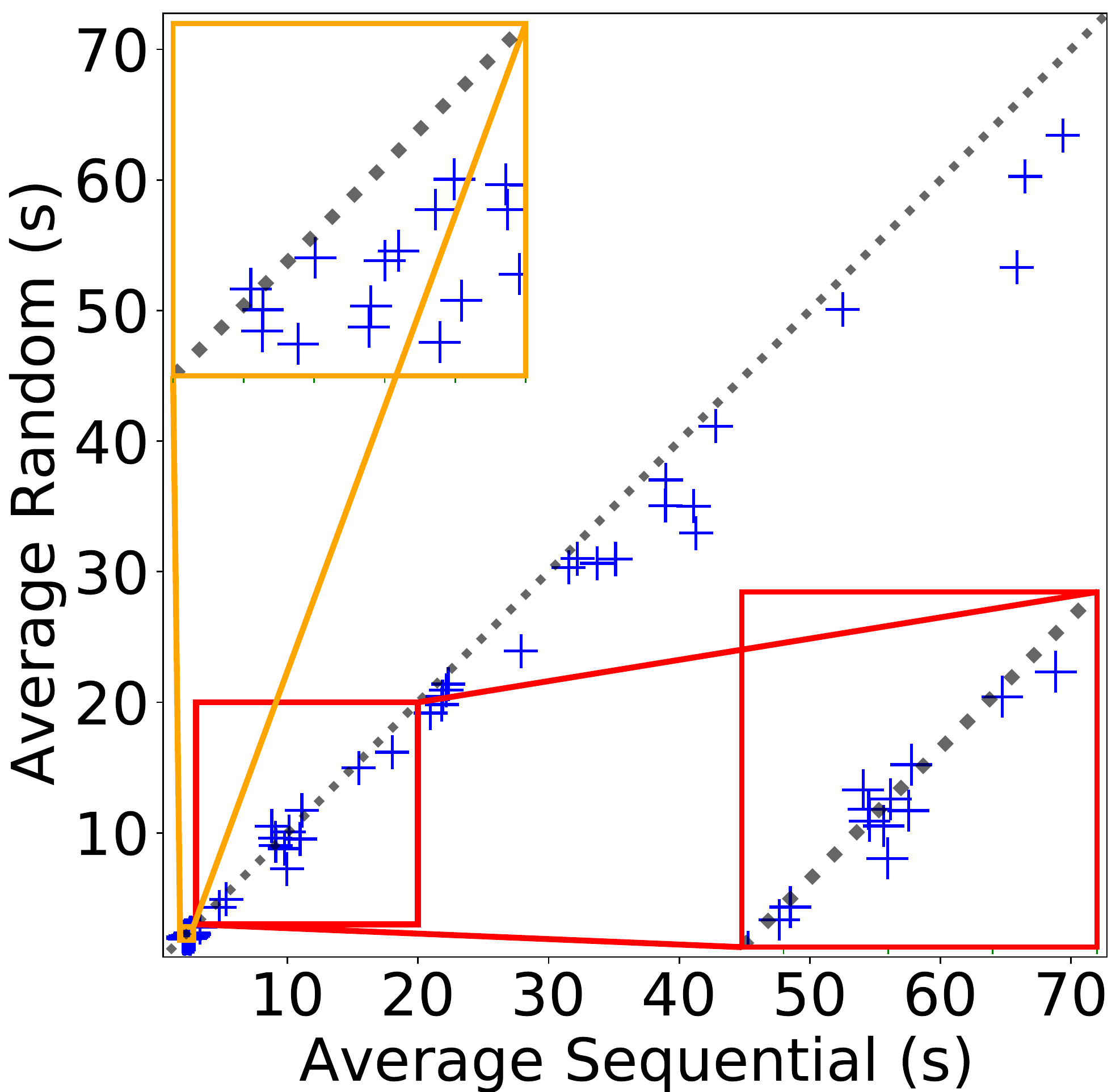}
  \end{minipage}
  \hfill
  \begin{minipage}[b]{0.49\columnwidth}
    \includegraphics[width=\columnwidth]{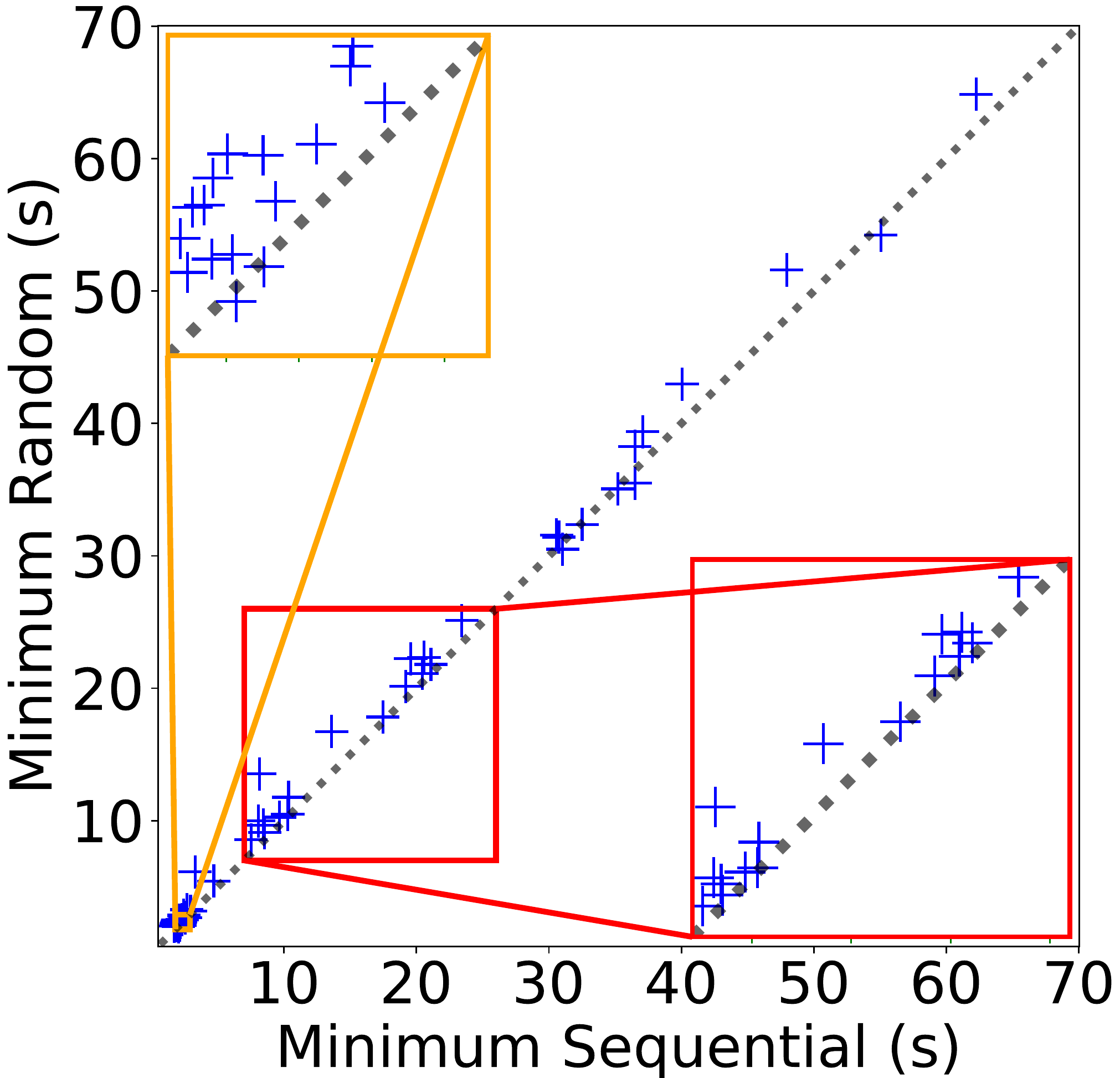}
  \end{minipage}
 \caption{{\textbf{ALL STATES $\backslash$ DL-S-POR $\backslash$ BUG: SINGLE INVOCATION}} \newline Time comparison between sequential and random strategies when {\lazy} computes all states, using clients in which only a single invocation is buggy.}
   \label{Fig:AllLess}
\end{figure}

\begin{figure}[!tbp]
  \centering
  \begin{minipage}[b]{0.49\columnwidth}
    \includegraphics[width=\columnwidth]{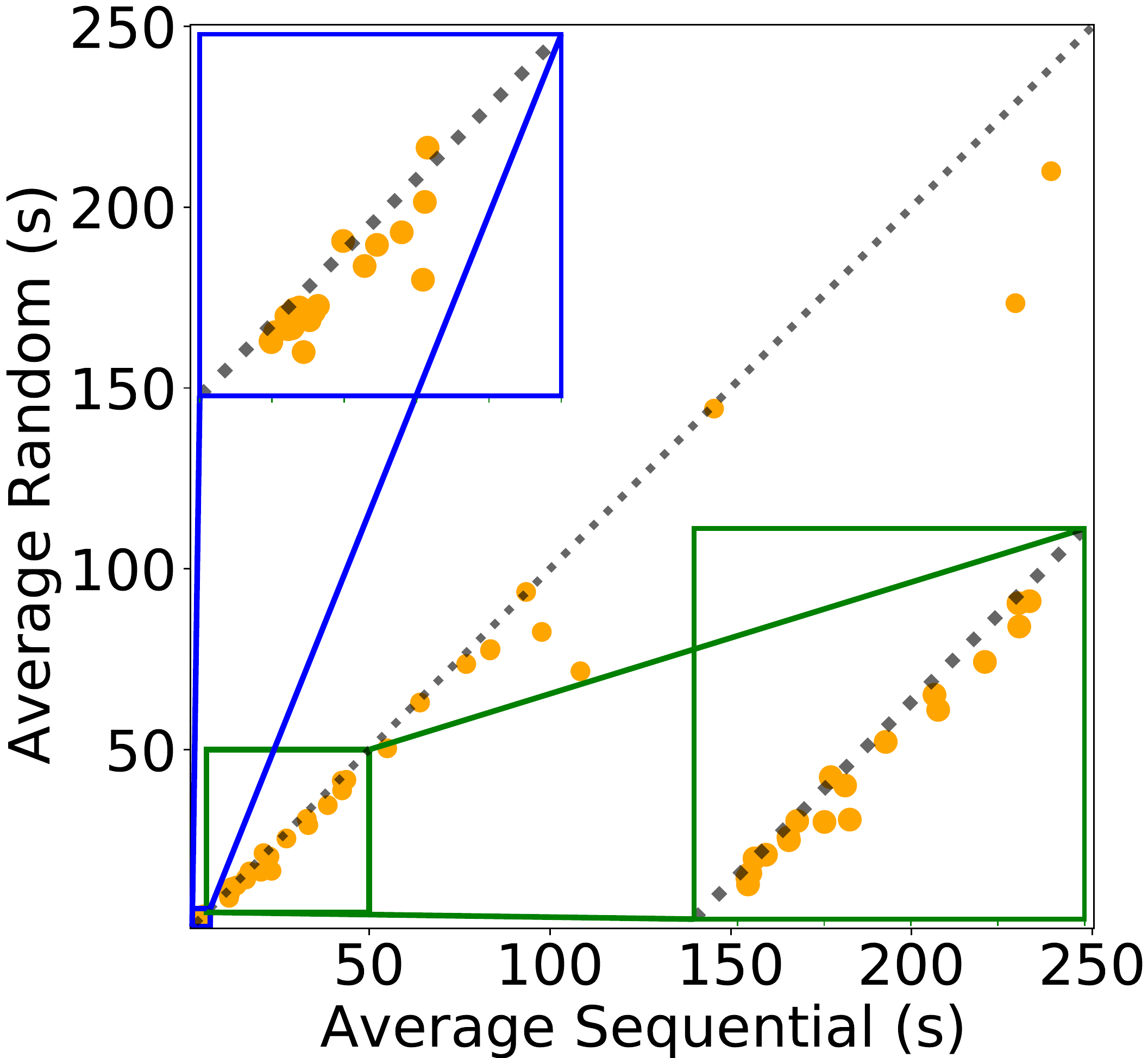}
  \end{minipage}
  \hfill
  \begin{minipage}[b]{0.49\columnwidth}
    \includegraphics[width=\columnwidth]{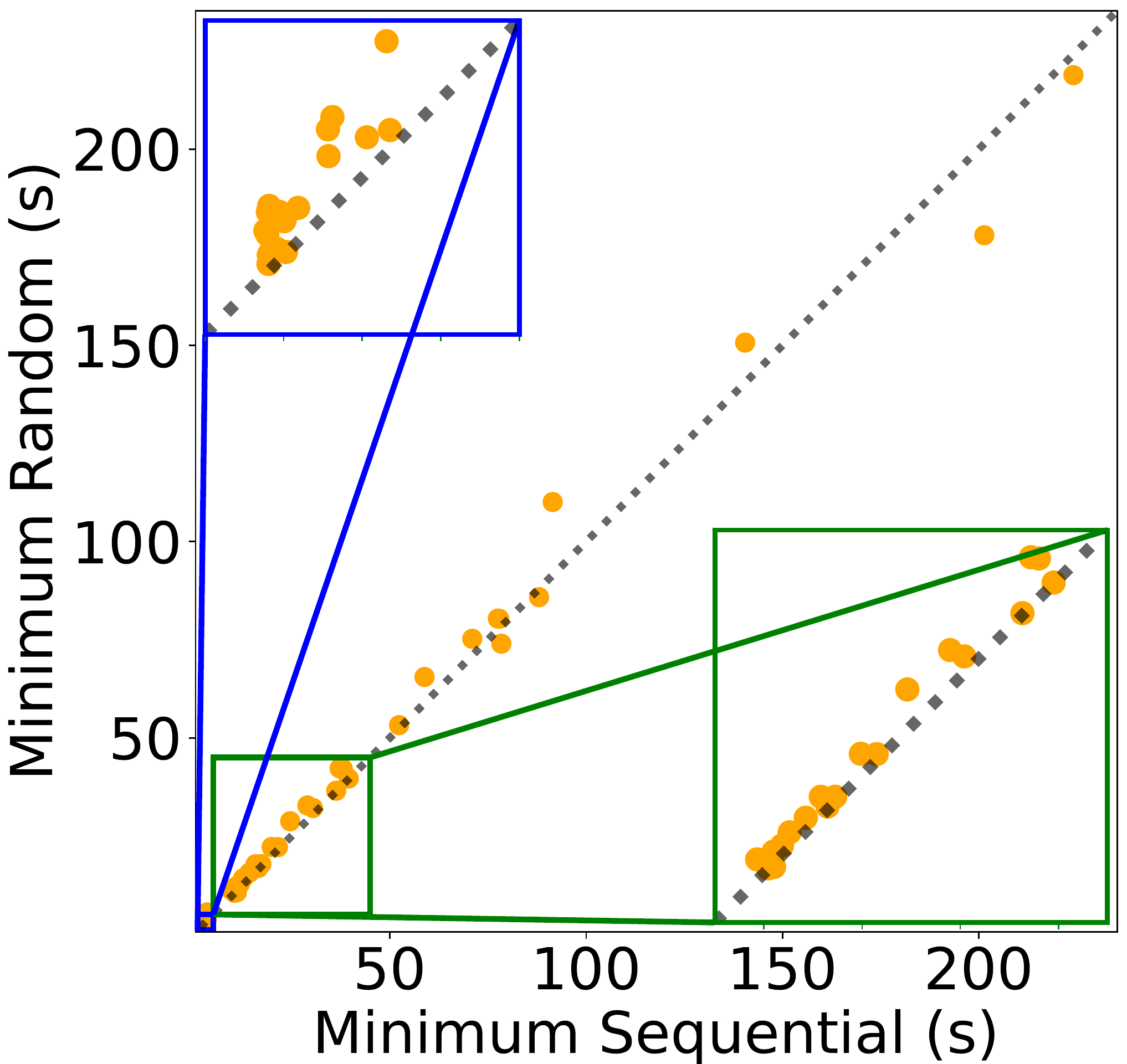}
  \end{minipage}
   \caption{{\textbf{ALL STATES $\backslash$ DL-S-POR $\backslash$ BUG: \textonehalf \ OF THE INVOCATIONS }} \newline Time comparison between sequential and random strategies when {\lazy} computes all states, using clients in which half of the invocations are buggy.}
  \label{Fig:AllMed}
\end{figure}

\begin{figure}[!tbp]
  \centering
  \begin{minipage}[b]{0.49\columnwidth}
    \includegraphics[width=\columnwidth]{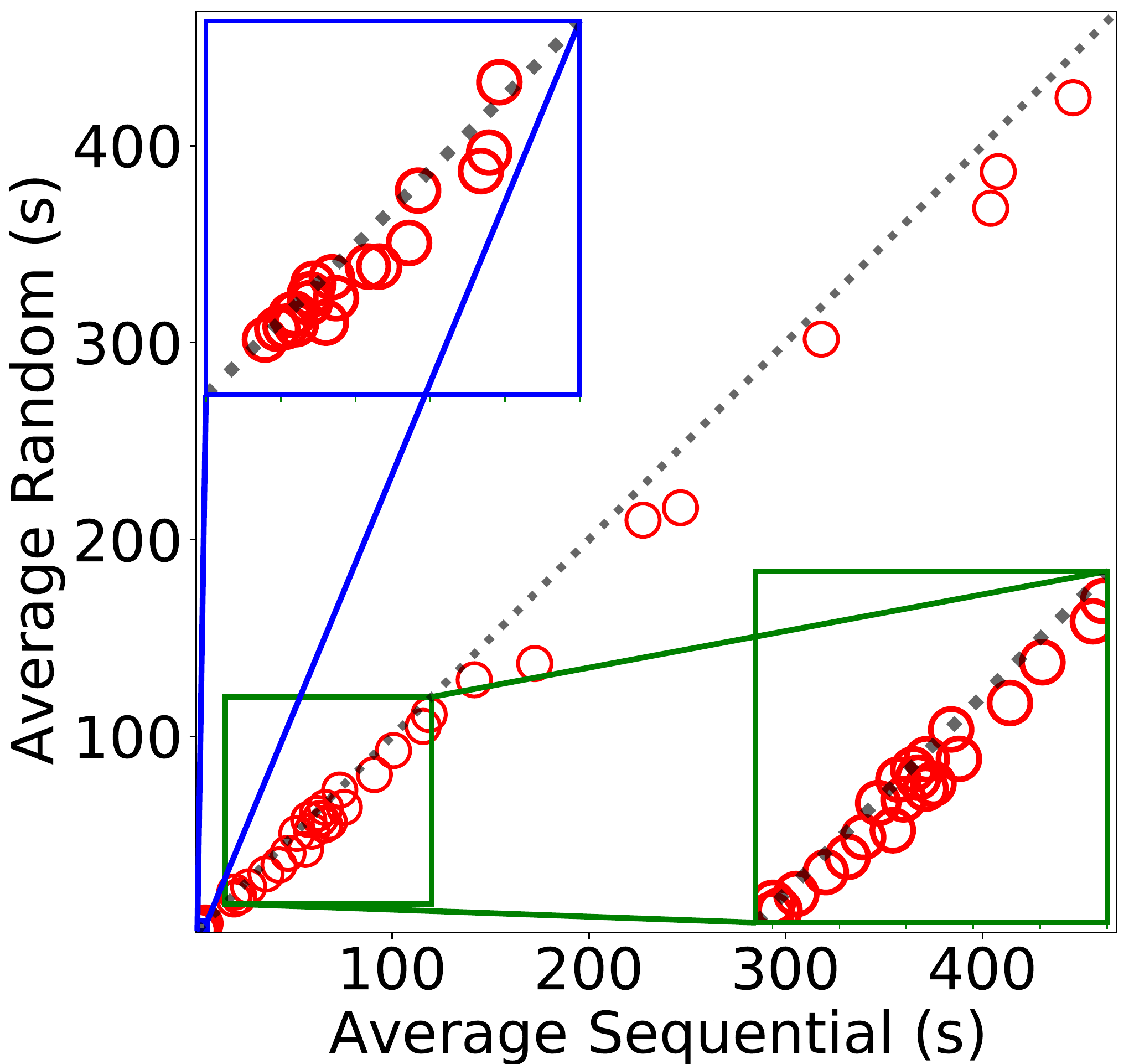}
  \end{minipage}
  \hfill
  \begin{minipage}[b]{0.49\columnwidth}
    \includegraphics[width=\columnwidth]{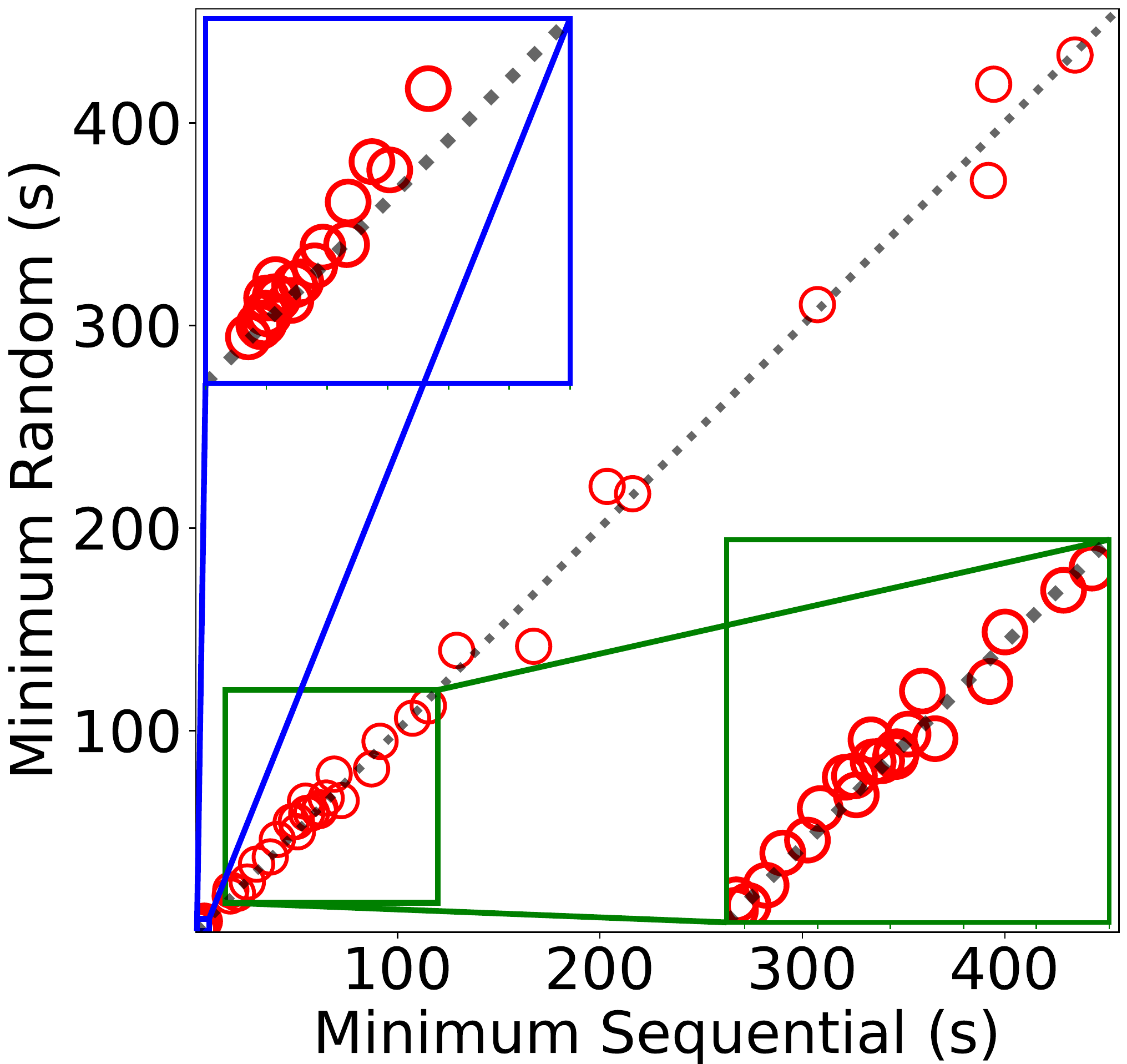}
  \end{minipage}
  \caption{{\textbf{ALL STATES $\backslash$ DL-S-POR $\backslash$ BUG: ALL INVOCATIONS}} \newline Time comparison between sequential and random strategies when {\lazy} computes all states, using clients in which all of the invocations are buggy.}
  \label{Fig:AllMore}
\end{figure}

\begin{figure}[!tbp]
  \centering
  \begin{minipage}[b]{0.49\columnwidth}
    \includegraphics[width=0.75\columnwidth]{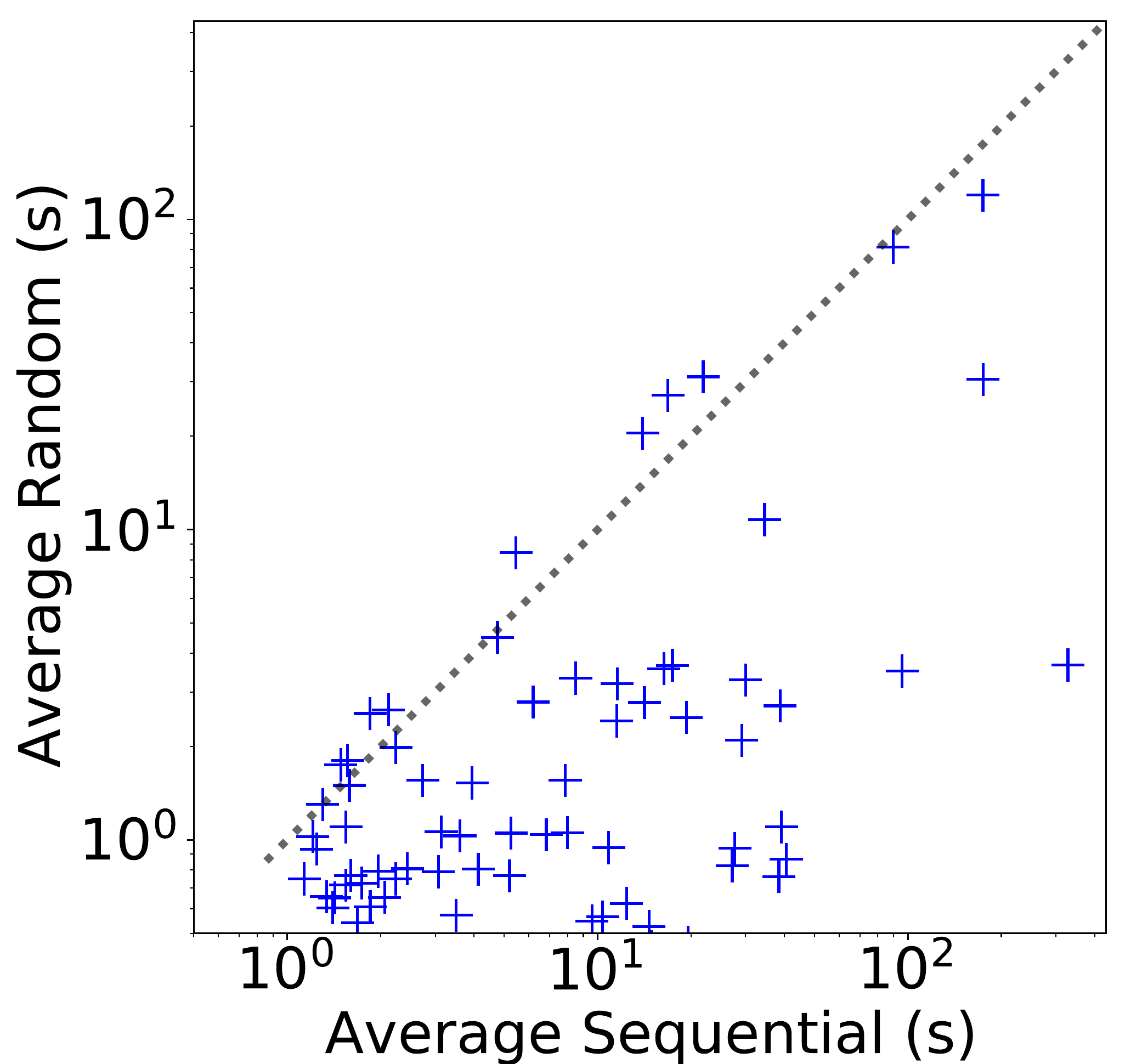}
  \end{minipage}
  \hfill
  \begin{minipage}[b]{0.49\columnwidth}
    \includegraphics[width=0.75\columnwidth]{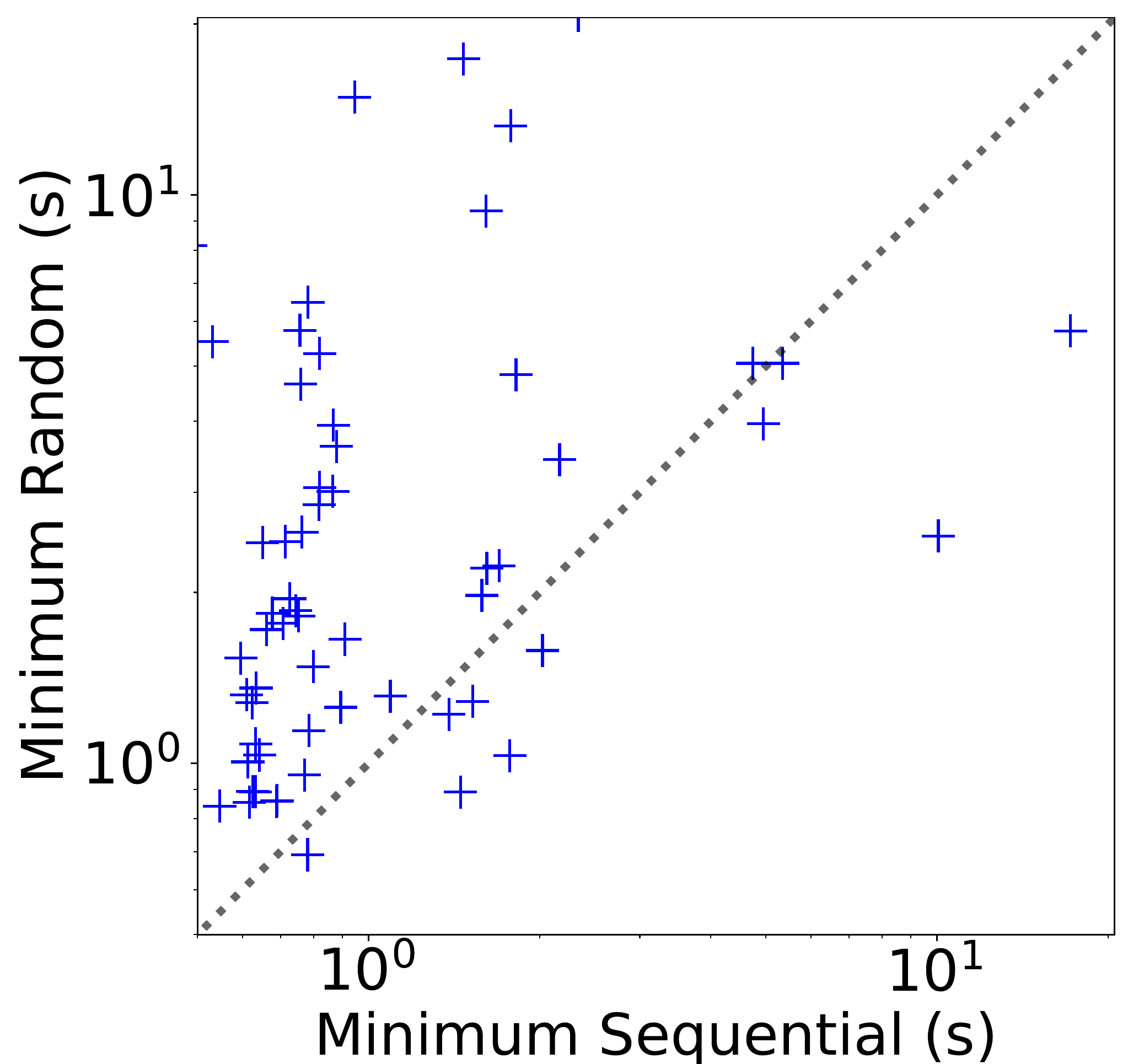}
  \end{minipage}
  \caption{{\textbf{FIRST ERROR $\backslash$ S-POR $\backslash$ BUG: SINGLE INVOCATION}} \newline Time comparison (log scale) between sequential and random strategies when {\ample} enumerates states only until the first error, using clients in which only a single invocation is buggy.}
  \label{Fig:FirstAmpleLess}
\end{figure}

\begin{figure}[!tbp]
  \centering
  \begin{minipage}[b]{0.49\columnwidth}
    \includegraphics[width=0.75\columnwidth]{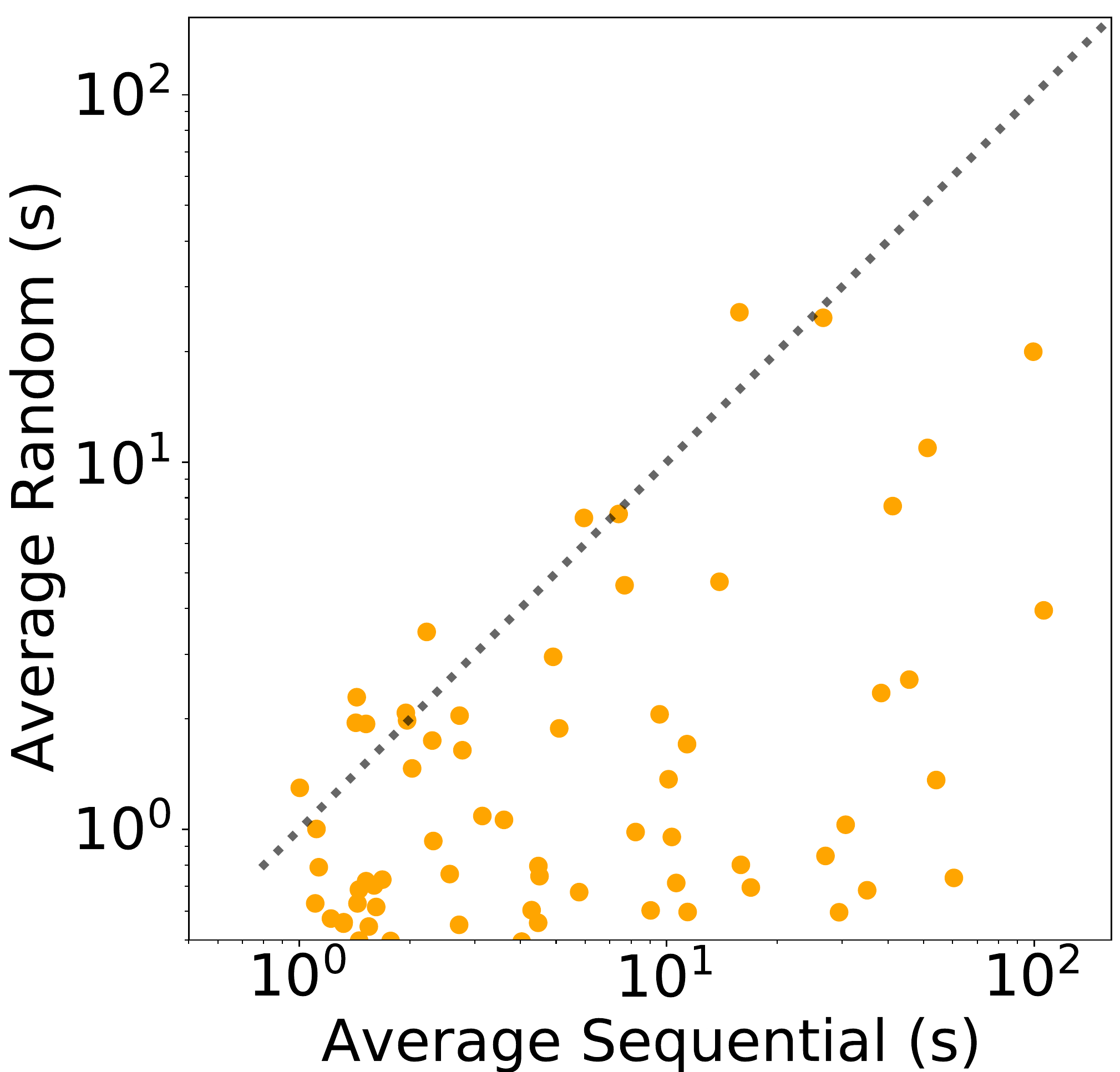}
  \end{minipage}
  \hfill
  \begin{minipage}[b]{0.49\columnwidth}
    \includegraphics[width=0.75\columnwidth]{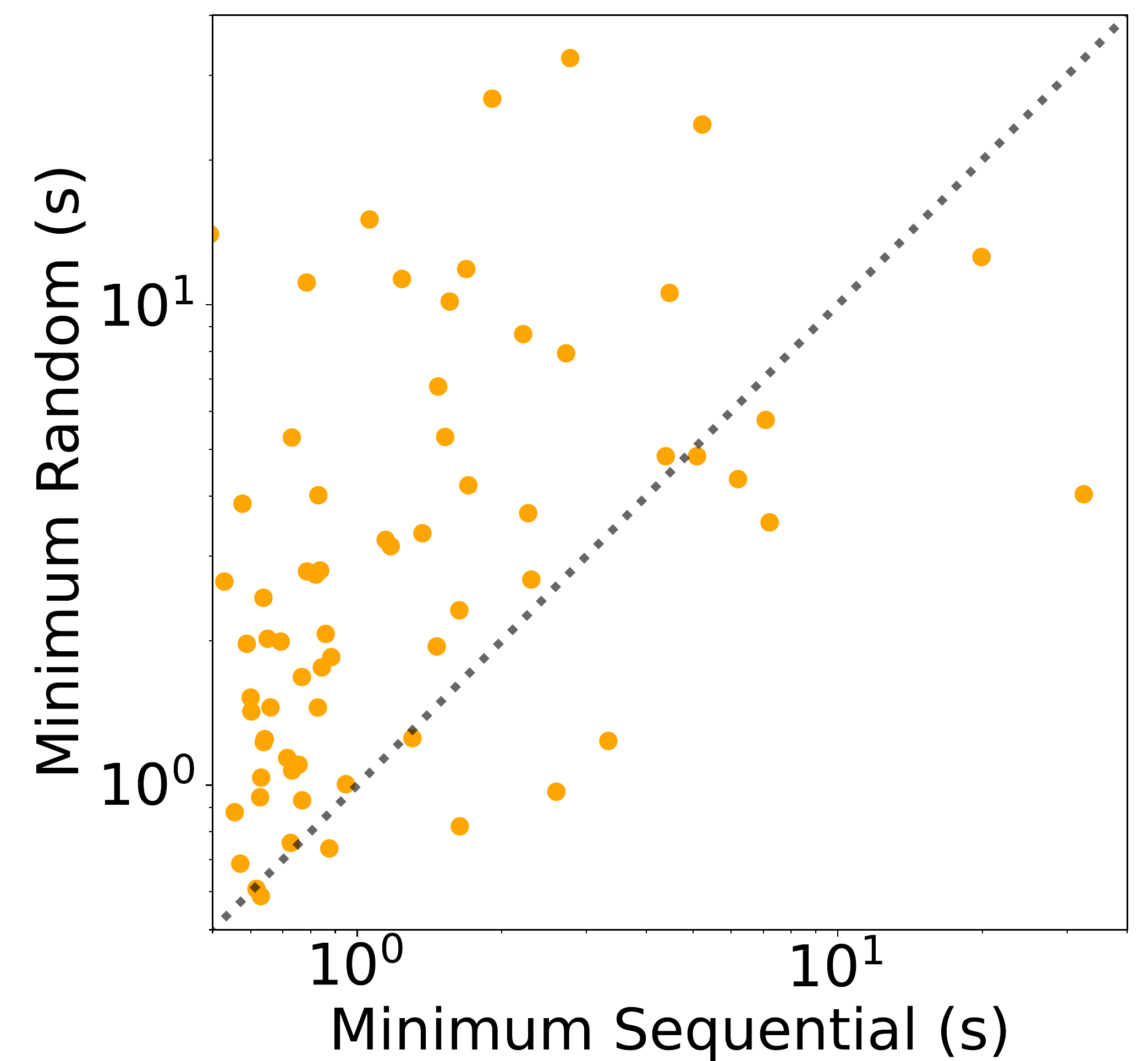}
  \end{minipage}
 \caption{{\textbf{FIRST ERROR $\backslash$ S-POR $\backslash$ BUG: \textonehalf \ OF THE INVOCATIONS}} \newline Time comparison (log scale) between sequential and random strategies when {\ample} enumerates states only until the first error, using clients in which half of the invocations are buggy.}
  \label{Fig:FirstAmpleMed}
\end{figure}

\begin{figure}[!tbp]
  \centering
  \begin{minipage}[b]{0.49\columnwidth}
    \includegraphics[width=0.75\columnwidth]{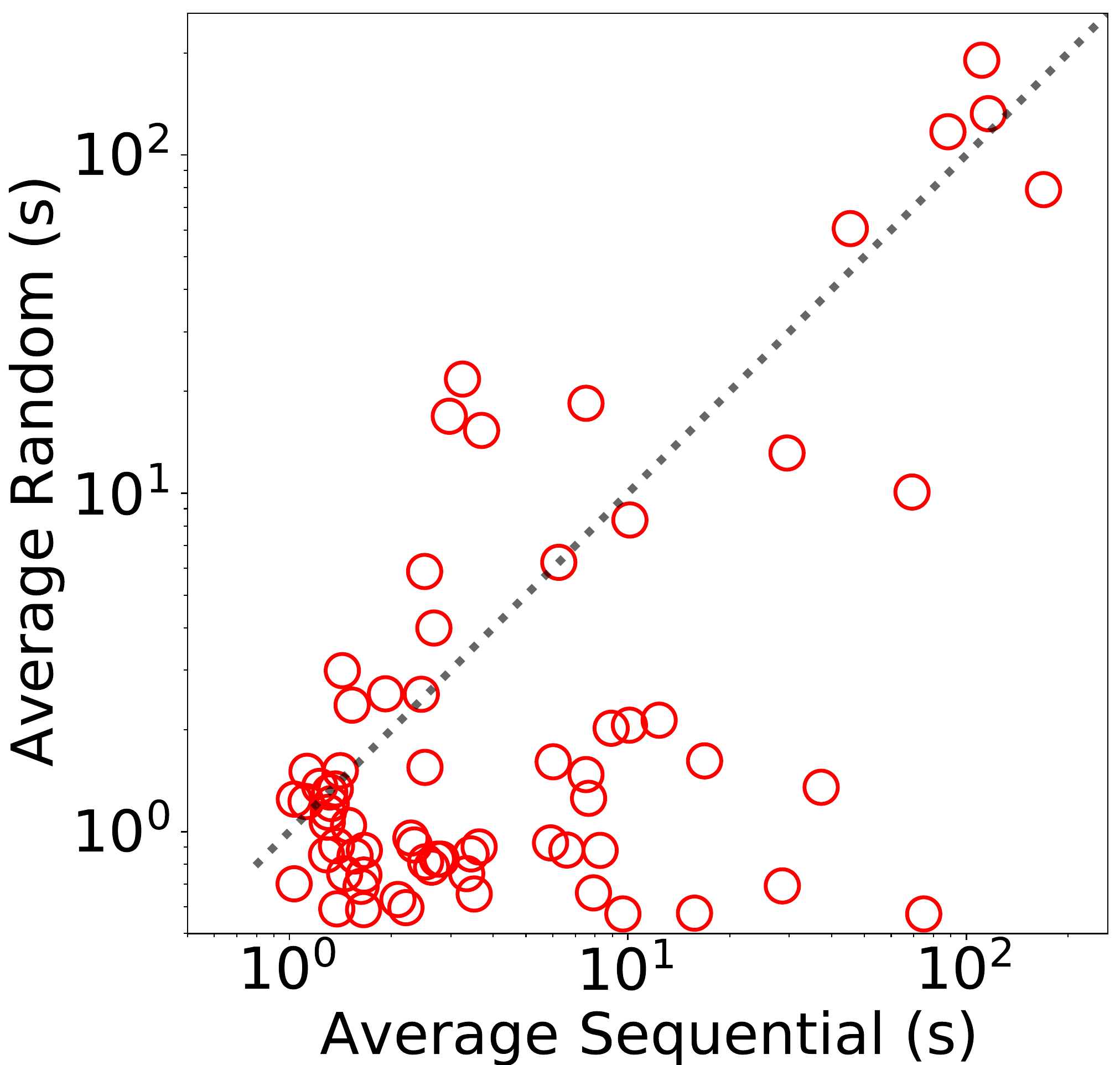}
  \end{minipage}
  \hfill
  \begin{minipage}[b]{0.49\columnwidth}
    \includegraphics[width=0.75\columnwidth]{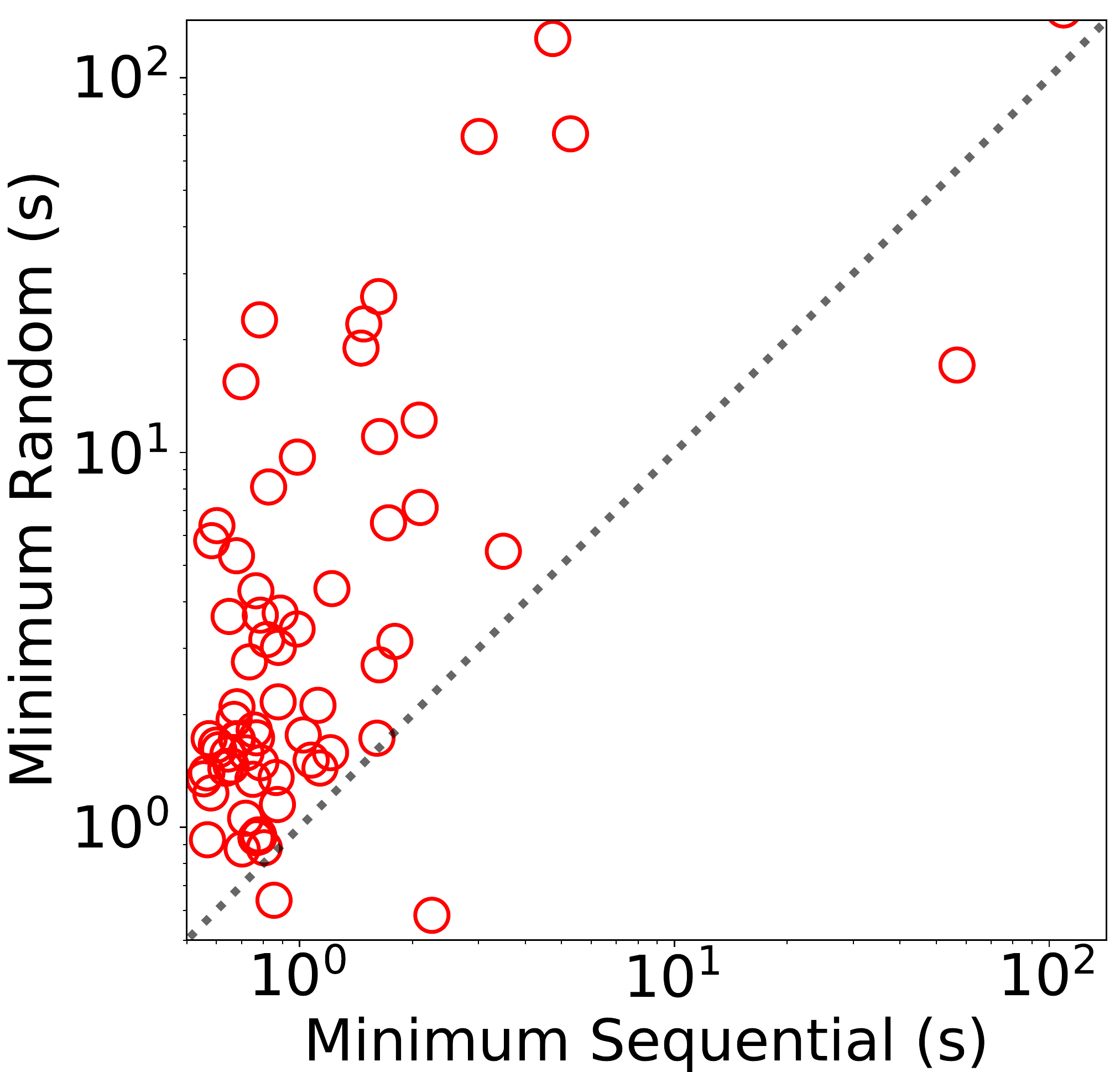}
  \end{minipage}
 \caption{{\textbf{FIRST ERROR $\backslash$ S-POR $\backslash$ BUG: ALL INVOCATIONS}} \newline Time comparison (log scale) between sequential and random strategies when {\ample} enumerates states only until the first error, using clients in which all of the invocations are buggy.}
  \label{Fig:FirstAmpleMore}
\end{figure}

\begin{figure}[!tbp]
  \centering
  \begin{minipage}[b]{0.49\columnwidth}
    \includegraphics[width=0.75\columnwidth]{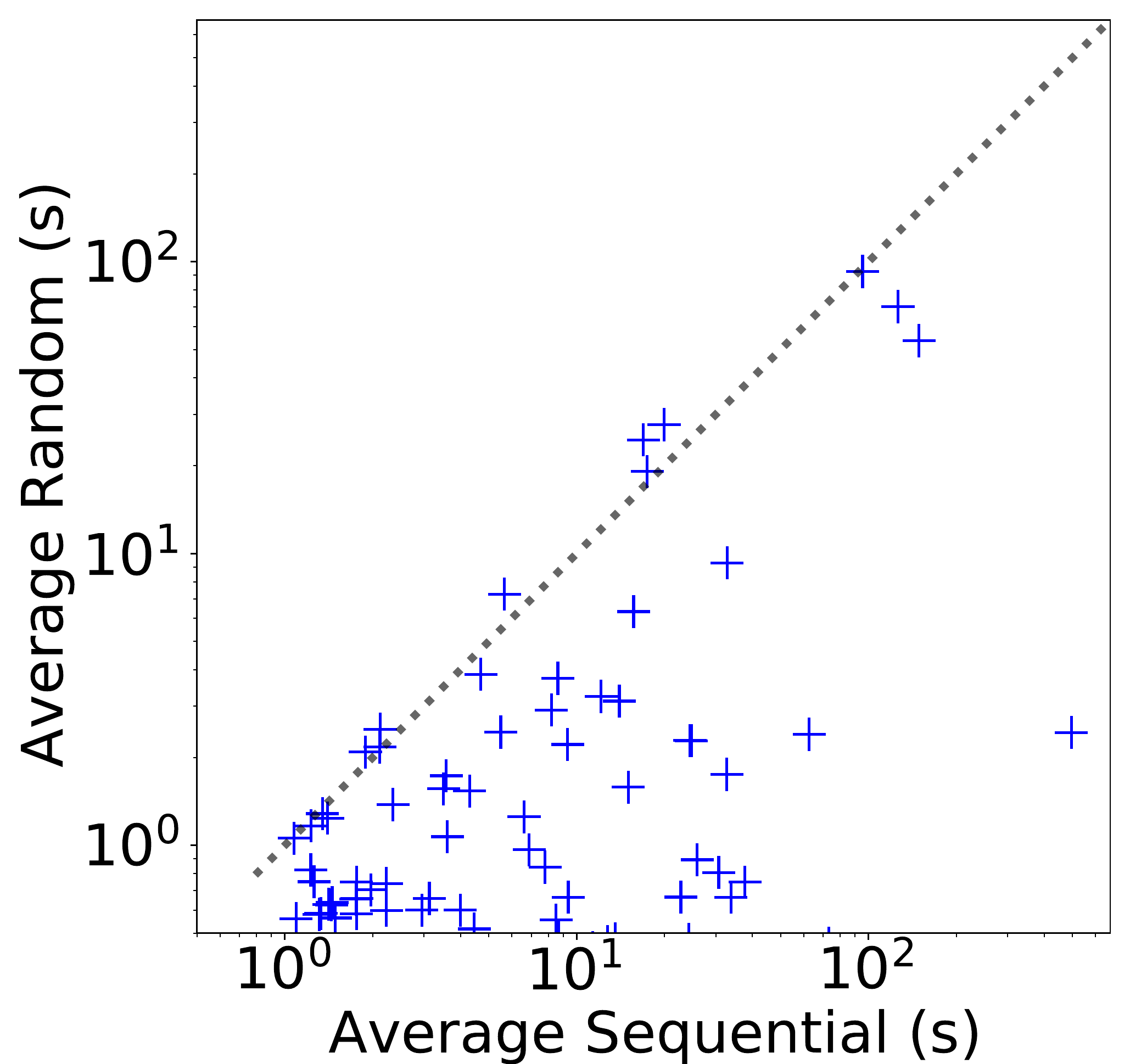}
  \end{minipage}
  \hfill
  \begin{minipage}[b]{0.49\columnwidth}
    \includegraphics[width=0.75\columnwidth]{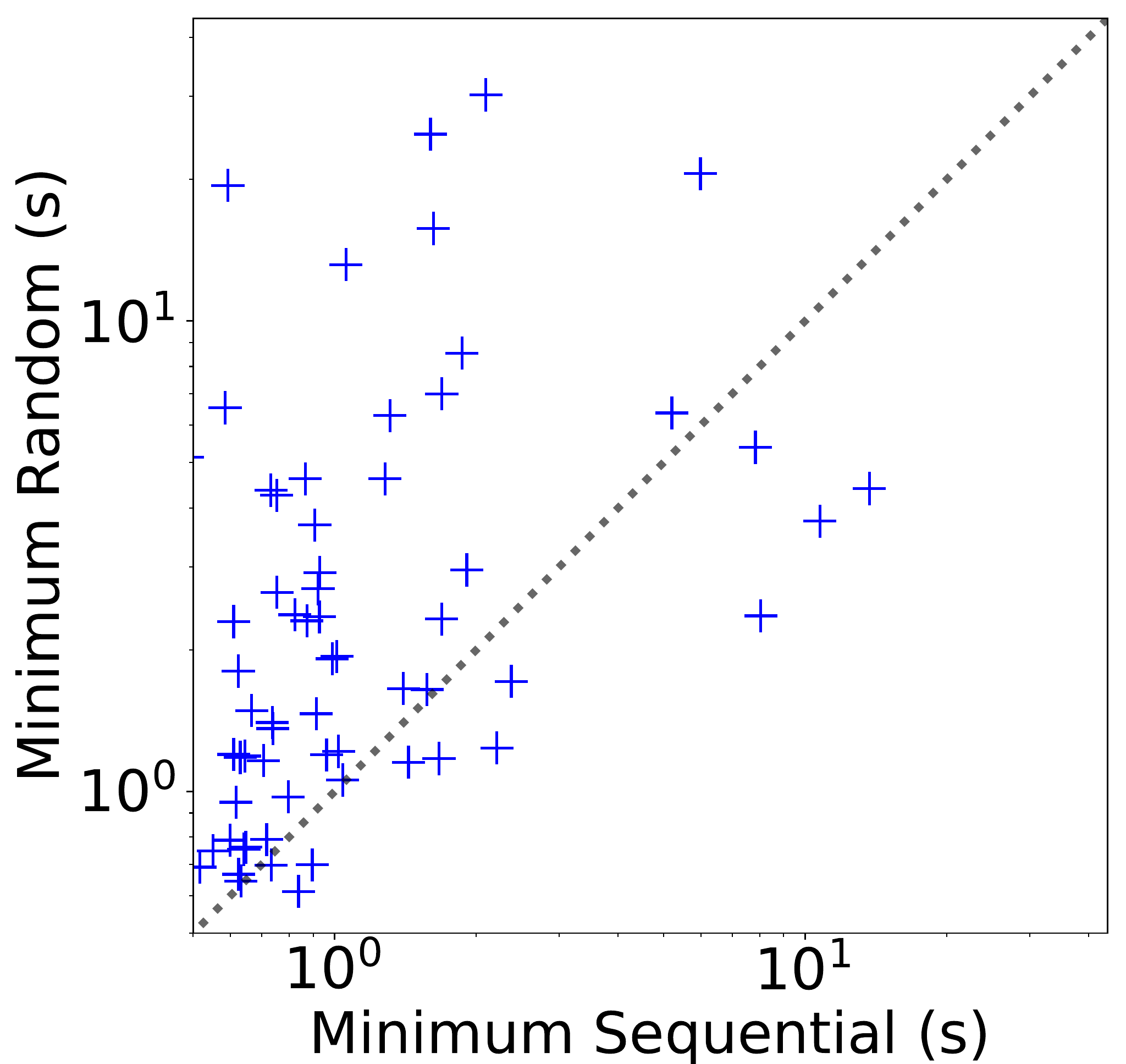}
  \end{minipage}
   \caption{{\textbf{FIRST ERROR $\backslash$ DL-S-POR $\backslash$ BUG: SINGLE INVOCATION}} \newline Time comparison (log scale) between sequential and random strategies when {\lazy} enumerates states only until the first error, using clients in which only a single invocation is buggy.}
  \label{Fig:FirstLazyLess}
\end{figure}

\begin{figure}[!tbp]
  \centering
  \begin{minipage}[b]{0.49\columnwidth}
    \includegraphics[width=0.75\columnwidth]{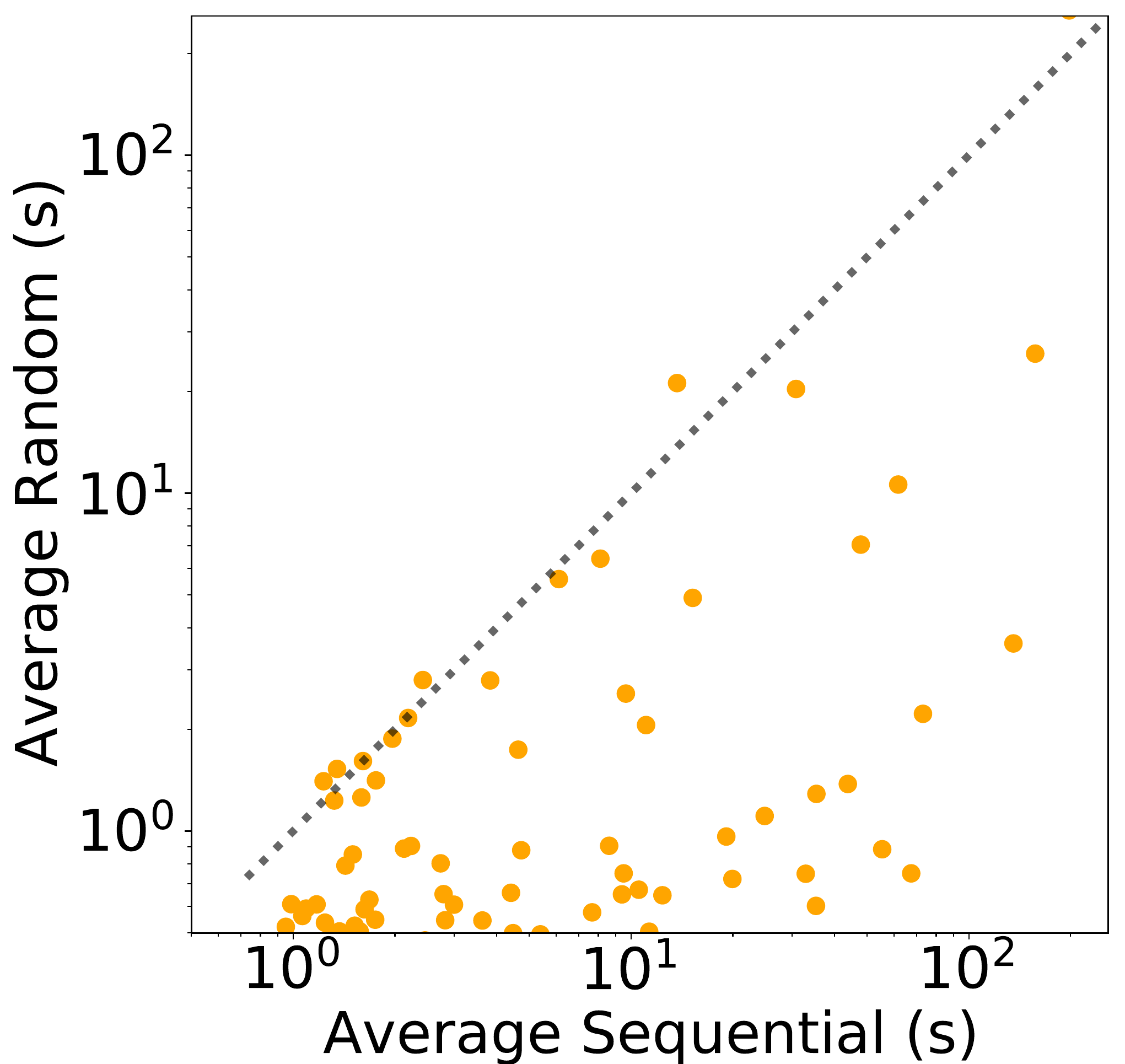}
  \end{minipage}
  \hfill
  \begin{minipage}[b]{0.49\columnwidth}
    \includegraphics[width=0.75\columnwidth]{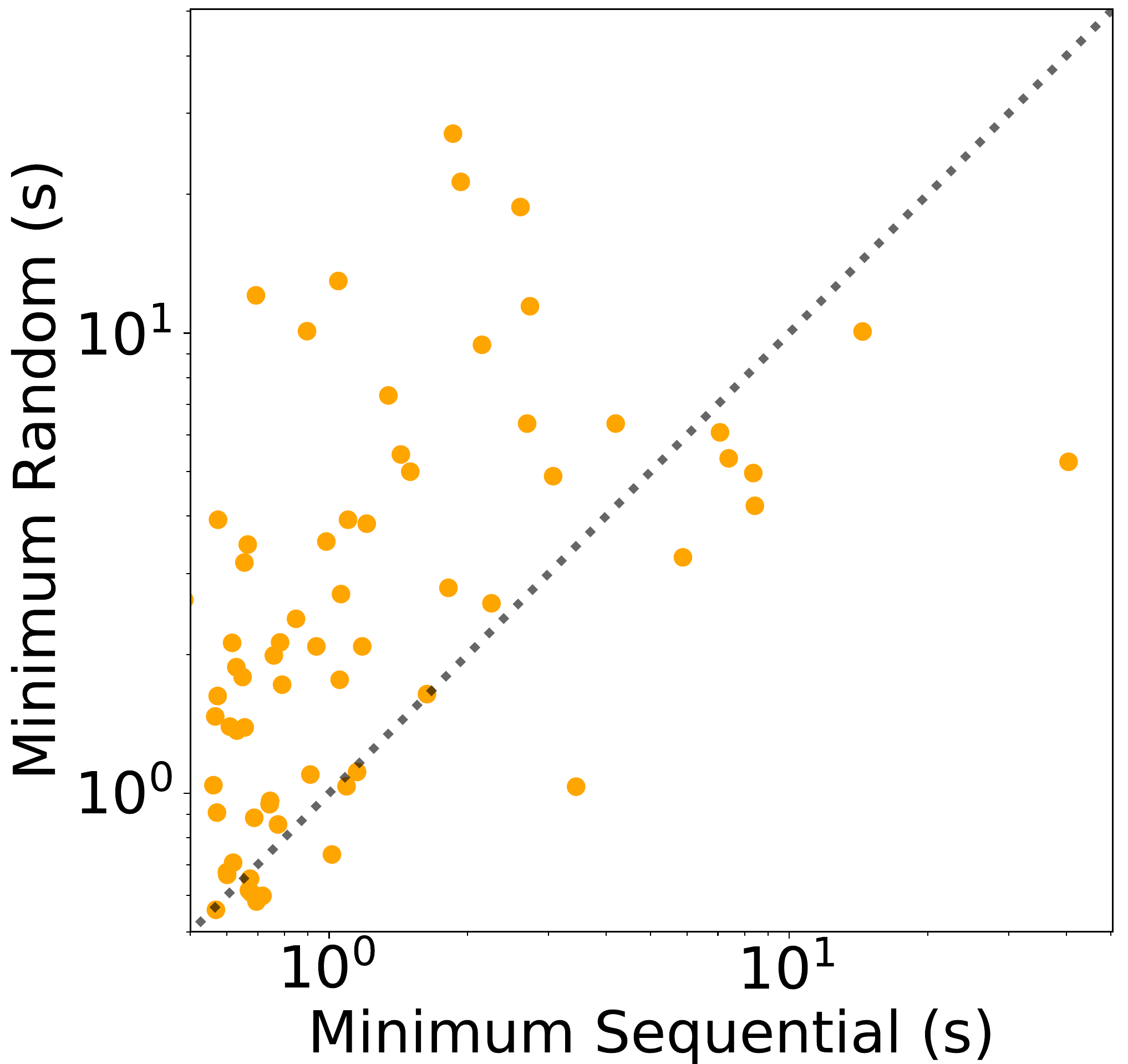}
  \end{minipage}
   \caption{{\textbf{FIRST ERROR $\backslash$ DL-S-POR $\backslash$ BUG: \textonehalf \ OF THE INVOCATIONS}} \newline Time comparison (log scale) between sequential and random strategies when {\lazy} enumerates states only until the first error, using clients in which half of the invocations are buggy.}
  \label{Fig:FirstLazyMed}
\end{figure}

\begin{figure}[!tbp]
  \centering
  \begin{minipage}[b]{0.49\columnwidth}
    \includegraphics[width=0.75\columnwidth]{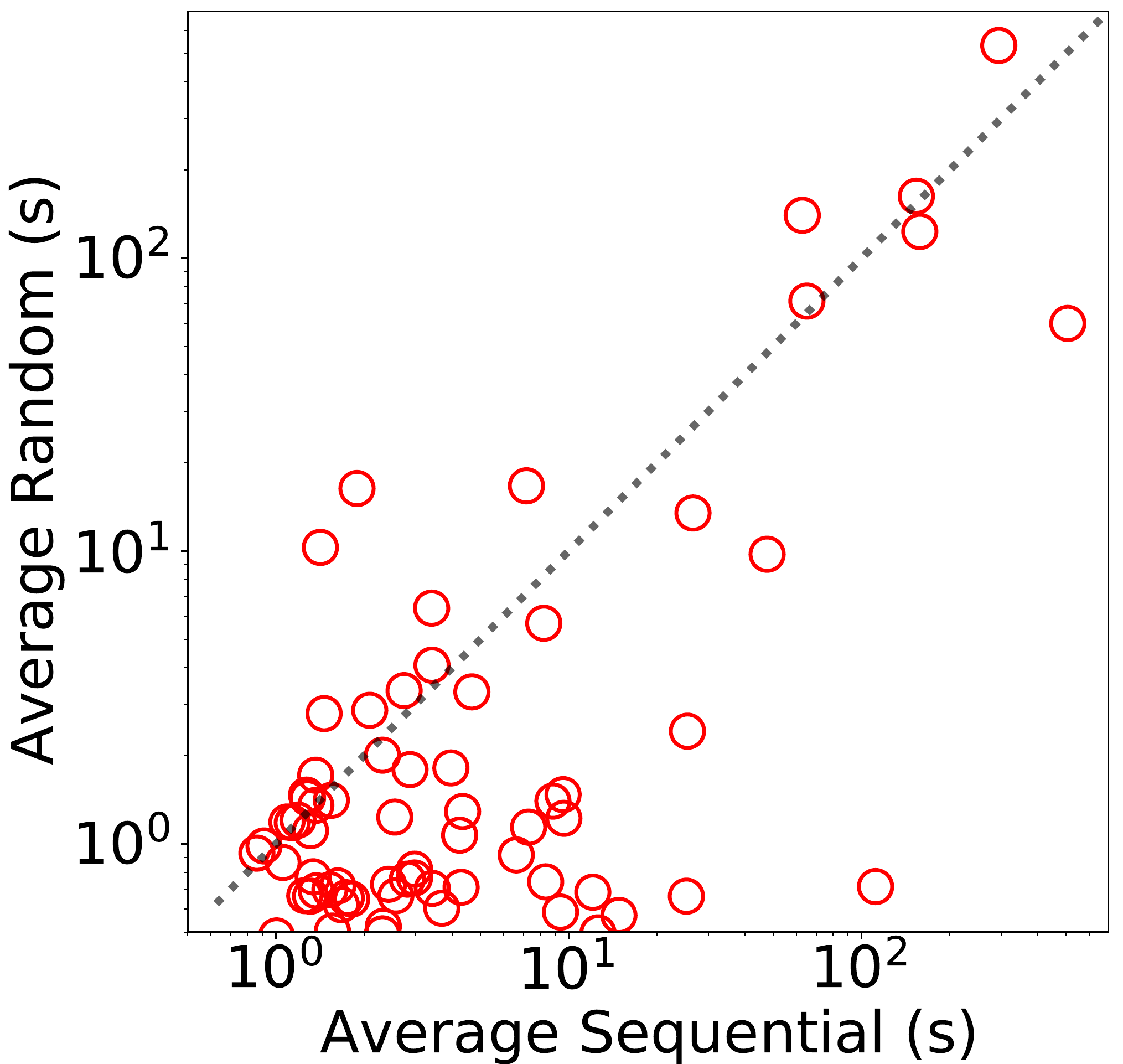}
  \end{minipage}
  \hfill
  \begin{minipage}[b]{0.49\columnwidth}
    \includegraphics[width=0.75\columnwidth]{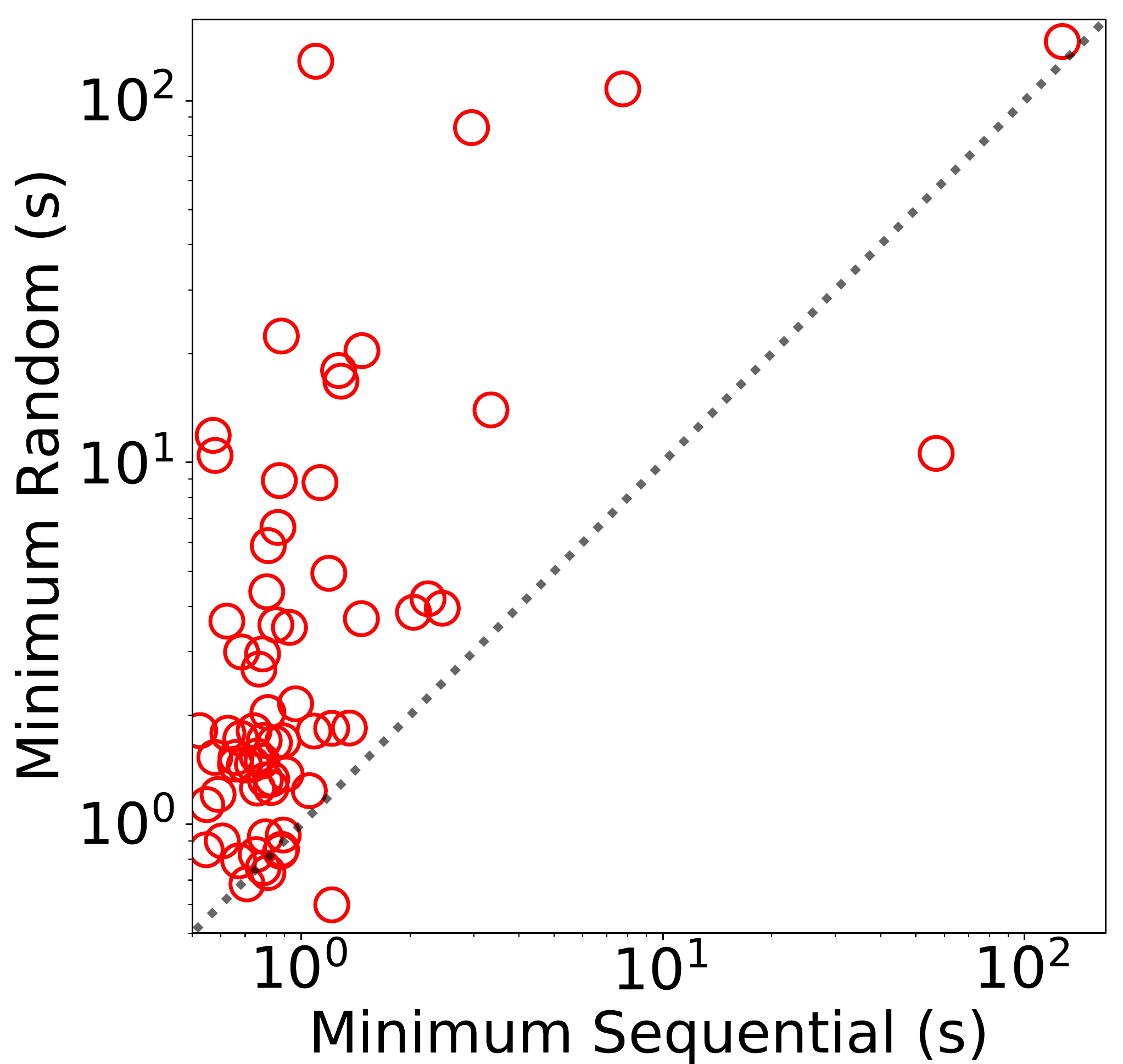}
  \end{minipage}
    \caption{{\textbf{FIRST ERROR $\backslash$ DL-S-POR $\backslash$ BUG: ALL INVOCATIONS}} \newline Time comparison (log scale) between sequential and random strategies when {\lazy} enumerates states only until the first error, using clients in which all of the invocations are buggy.}
  \label{Fig:FirstLazyMore}
\end{figure}

We present more results obtained from the experiments for pair-wise comparison between the random and sequential enumeration strategies. We report the average and minimum time over the different instances as follows: 

\begin{itemize}
\item The first set of experiments are for computing all reachable states, only with {\lazy}. For {\ample}, the enumeration strategy is not important since there is no dynamic computation of backtrack sets. We present separate figures for each different class of clients based on the number of invocations to methods that lead to bugs: (1) none of the invocations, (Figure~\ref{Fig:AllCorrect}), (2) just a single invocation (Figure~\ref{Fig:AllLess}) (3) half of the invocations (Figure~\ref{Fig:AllMed}) and (4) all of the invocations (Figure~\ref{Fig:AllMore}). 
\item The second set of experiments are for computing reachable states only until the first error, both with {\ample} (Figures from \ref{Fig:FirstAmpleLess} to \ref{Fig:FirstAmpleMore}) and {\lazy} (Figures from \ref{Fig:FirstLazyLess} to \ref{Fig:FirstLazyMore}). As the purpose is to find the first bug, clients in which none of the invocations are buggy, are not included in this set of experiments. Figures for the rest of the classes of clients are represented with the same order and visualization as in the first set. Differently from the first set of figures, the figures in second set are in logarithmic scale.
\end{itemize}

\end{document}